\documentclass[twocolumn]{aastex631}

\graphicspath{{./}{figures/}}

\usepackage{amsmath}

\begin{document}

\title{Estimating Initial Mass of Gaia-Enceladus Dwarf Galaxy with Chemical Evolution Model}

\author[0000-0002-0435-4493]{Olcay PLEVNE}
\affiliation{Faculty of Science, Department of Astronomy and Space Sciences, Istanbul University, Istanbul, Turkey}

\author[0000-0002-9993-7244]{Furkan AKBABA}
\affiliation{Institute of Graduate Studies in Science, Istanbul University, Istanbul, Turkey}

\newcommand{\danny}[1]{\textcolor{blue}{\textbf{DH}: #1}}



\begin{abstract}
This work investigates the initial mass and chemical evolution history of the Gaia-Enceladus dwarf galaxy. We combine spectroscopic data from APOGEE with astrometric data from Gaia DR3 to identify Gaia-Enceladus candidate stars via a machine-learning pipeline using t-SNE and HDBSCAN. By focusing on kinematic and chemical parameters, especially $\mathrm{[Fe/H]}$, $\mathrm{[Mg/Fe]}$, $\mathrm{[Al/Fe]}$, and $\mathrm{[Mn/Fe]}$, we uncover a population of metal-poor, high-eccentricity stars that align with literature criteria for Gaia-Enceladus debris. We then apply the \textit{OMEGA+} chemical evolution model, incorporating MCMC fitting of the observed abundance trends in the $\mathrm{[Mg/Fe]\times[Fe/H]}$ plane. Our best-fitting model indicates a gas mass of $4.93_{-0.72}^{+0.32}\times10^9\,{M_{\odot}}$ for Gaia-Enceladus, placing it at the higher end of previously suggested mass ranges. The model scenario suggests a short star formation timescale, substantial outflows, and a rapid build-up of metals mainly driven by core-collapse supernovae, with a lesser contribution from Type~Ia supernovae. Comparison with observational data in other chemical planes (e.g., $\mathrm{[Mg/Mn]\times[Al/Fe]}$) supports this scenario, emphasizing a distinct evolution path relative to the Milky Way. Additionally, our results provide indirect evidence that star formation in Gaia-Enceladus likely ceased within the first 4 Gyr, consistent with earlier inferences of an early merger event. These findings highlight the power of chemical evolution modeling in reconstructing the origin and mass of ancient accreted systems. Overall, we show that Gaia-Enceladus, through a rapid star formation and strong outflows, contributed a significant fraction of the metal-poor stellar halo of the Milky Way.
\end{abstract}



\section{Introduction} \label{sec:intro}

Milky Way has attained its present-day structure through a combination of internal processes, continuous star formation, and external interactions, such as mergers with smaller systems over billions of years \citep{Freeman2002}. Over the last decade, our understanding of the Milky Way’s assembly history has significantly expanded, thanks in large part to extensive spectroscopic surveys (e.g., RAVE \citep{Steinmetzetal2006}, GALAH \citep{Galah}, APOGEE \citep{Apogee} and H3 \citep{H3survey}) and revolutionary astrometric data from Gaia \citep{Gaia16}. Each of these projects offers unique insights into elemental abundances, kinematics, and stellar ages across extensive Galactic volumes, thus enabling the differentiation of accreted versus in situ stellar components. These datasets enable the investigation of the chemical, kinematic, and photometric properties of millions of stars, thereby allowing for a more in-depth exploration of the Milky Way’s past merger events and their present-day signatures.

One of the most notable examples of such merger events is the discovery and characterization of the Gaia–Enceladus (also known as Gaia-Sausage/Enceladus) merger event \citep{Belokurov2018, Helmi2018, Haywood2018, Mackereth2019}. This ancient dwarf galaxy appears to dominate the inner halo of the Milky Way and is thought to have contributed a large fraction of the halo stars we observe today, particularly those with high orbital eccentricities and low metallicities ($e_p > 0.7,\,\mathrm{[Fe/H]} < -1~\mathrm{dex}$, \citet{Helmi2018}). However, its extended kinematic and chemical properties suggest that some stars previously associated with Gaia–Enceladus may have originated from other progenitors or even in-situ populations \citep{Feuillet2021, Horta2023}. On the other hand, recent studies imply that some dynamically or chemically distinct ex-situ substructures—such as Arjuna, I’itoi, or Sequoia—may actually trace back to the Gaia–Enceladus progenitor, reflecting the complex and overlapping signatures of early merger events \citep{Donlon2023, Mori2024, Nissen2024}. In fact, certain prominent phase-space substructures—such as Wukong, Arjuna, I’itoi, and Sequoia—have been proposed as possible fragments of the Gaia–Enceladus merger \citep[e.g.,][]{Mackereth2019, Naidu2020, Horta2022, Carrillo2023}. Clarifying which stars belong to Gaia–Enceladus and distinguishing them from other accreted or dynamically heated populations is therefore critical for understanding its role in the Milky Way’s evolution and assembling history.

Understanding the chemical evolution of Gaia–Enceladus is crucial for several reasons. First, by determining the star formation history, infall/outflow rates, and nucleosynthetic signatures of this dwarf galaxy, we can gain insight into the broader elemental cycle within the Milky Way. Second, Gaia–Enceladus offers a unique perspective on how early accretion events shape the structure and metallicity gradients of the Galactic halo and possibly the inner disk. Third, differentiating its stellar debris from other accreted populations—such as the Helmi streams or Sequoia or Sagittarius—helps clarify how multiple merger episodes influenced the evolution of the Milky Way’s bulk stellar populations over cosmic time.

However, separating Gaia–Enceladus—the largest known merger event in the Milky Way’s history \citep{Helmi2018}—from other halo populations or the Galactic field is not always straightforward. Conventional selection methods, based on chemical boundaries or simple orbital cuts, often suffer from contamination due to overlapping parameter distributions. Moreover, the presence of other minor merger remnants such as the Helmi streams \citep{Helmi_Streams}, Sequoia \citep{Barba2019, Matsuno2019, Myeong2019}, and Thamnos \citep{Koppelman2019} can mimic certain signatures once believed to be exclusive to Gaia–Enceladus. To address these issues, multi-dimensional approaches that combine detailed abundance ratios (e.g., $\mathrm{[Fe/H], [Mg/Fe], [Al/Fe], [Mn/Fe]}$) \citep{Hawkins2015, Das2020, Horta2023} with dynamical parameters (e.g., $E$, $L_z$) \citep{Belokurov2024} are required. Therefore, advanced analytical techniques are needed to evaluate these parameters jointly. In this context, machine learning tools—particularly unsupervised clustering algorithms—have proven especially effective in distinguishing halo substructures and mitigating contamination caused by subtle overlaps \citep{Carrillo2022}.

In this paper, we employ a combined spectroscopic and astrometric dataset from APOGEE DR17 \citep{ApogeeDR17} and Gaia DR3 \citep{GaiaDR3} to refine the selection of Gaia–Enceladus stars using a dimensionality-reduction and clustering pipeline. By incorporating a broader context of halo substructures, we aim to improve the identification of Gaia-Enceladus stars and better constrain its pre-merger stellar mass. We then integrate our selected stars into a chemical evolution framework, \textit{OMEGA+} \citep{Cote2017, Cote2018}, in order to reconstruct the evolutionary path of Gaia–Enceladus prior to its accretion.

Section~\ref{sec:data_selection} describes the dataset and the quality cuts applied to minimize observational uncertainties. In Section~\ref{sec:ml_pipeline}, we outline our machine learning pipeline used to isolate Gaia–Enceladus candidates, supported by diagnostic plots in various orbital and abundance planes. Section~\ref{sec:humanization} presents our cluster assignments, discusses the contamination sources, and evaluates the success of our method in capturing the bulk of Gaia–Enceladus debris. In Section~\ref{sec:chemev_model}, we demonstrate how these stars inform our chemical evolution model and subsequently revise estimates of Gaia–Enceladus’s progenitor properties. Finally, Section~\ref{sec:conclusion} provides a summary of our key findings and suggests directions for future work.

\begin{figure*}
    \centering
    \gridline{\fig{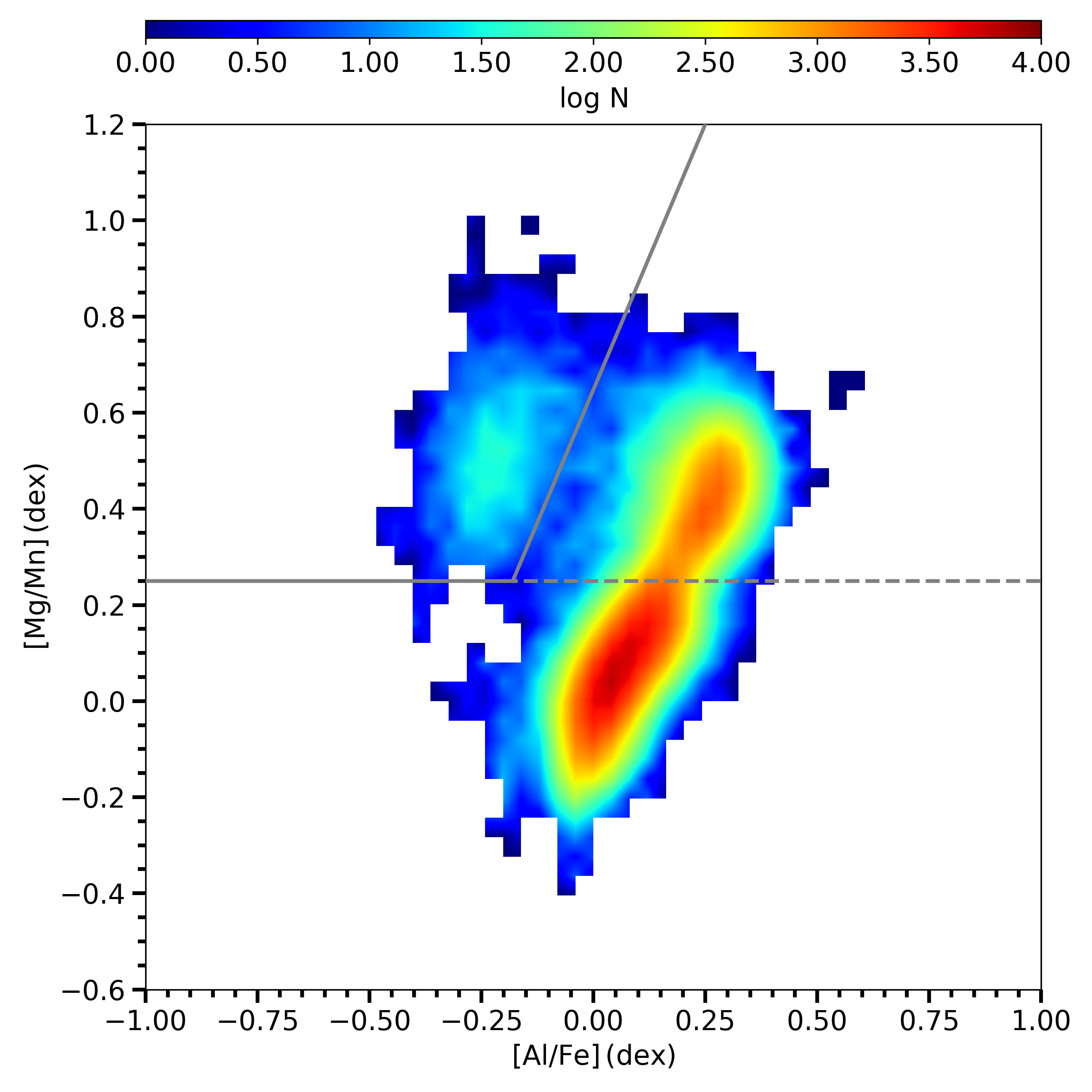}{0.30\textwidth}{}
          \fig{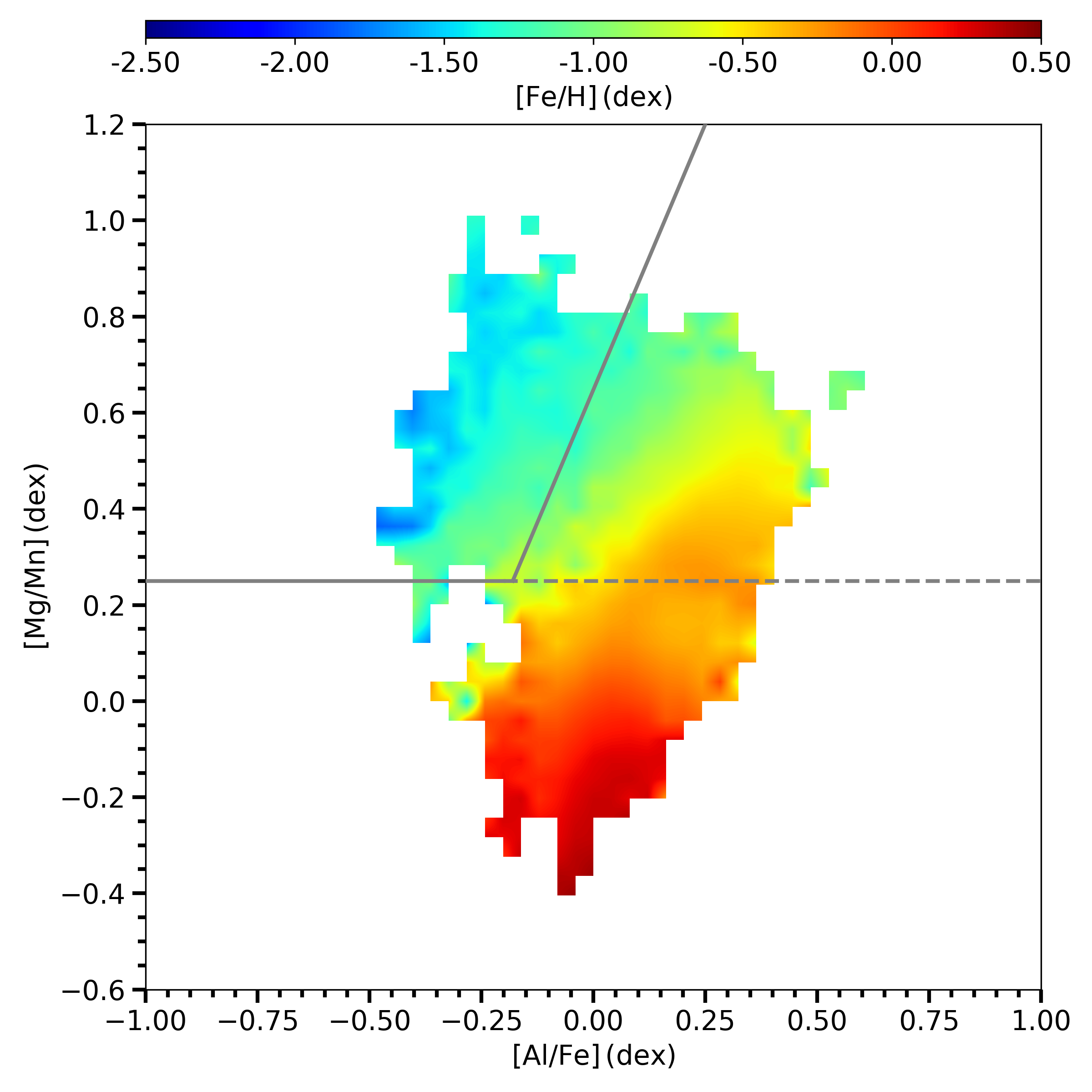}{0.30\textwidth}{}
          \fig{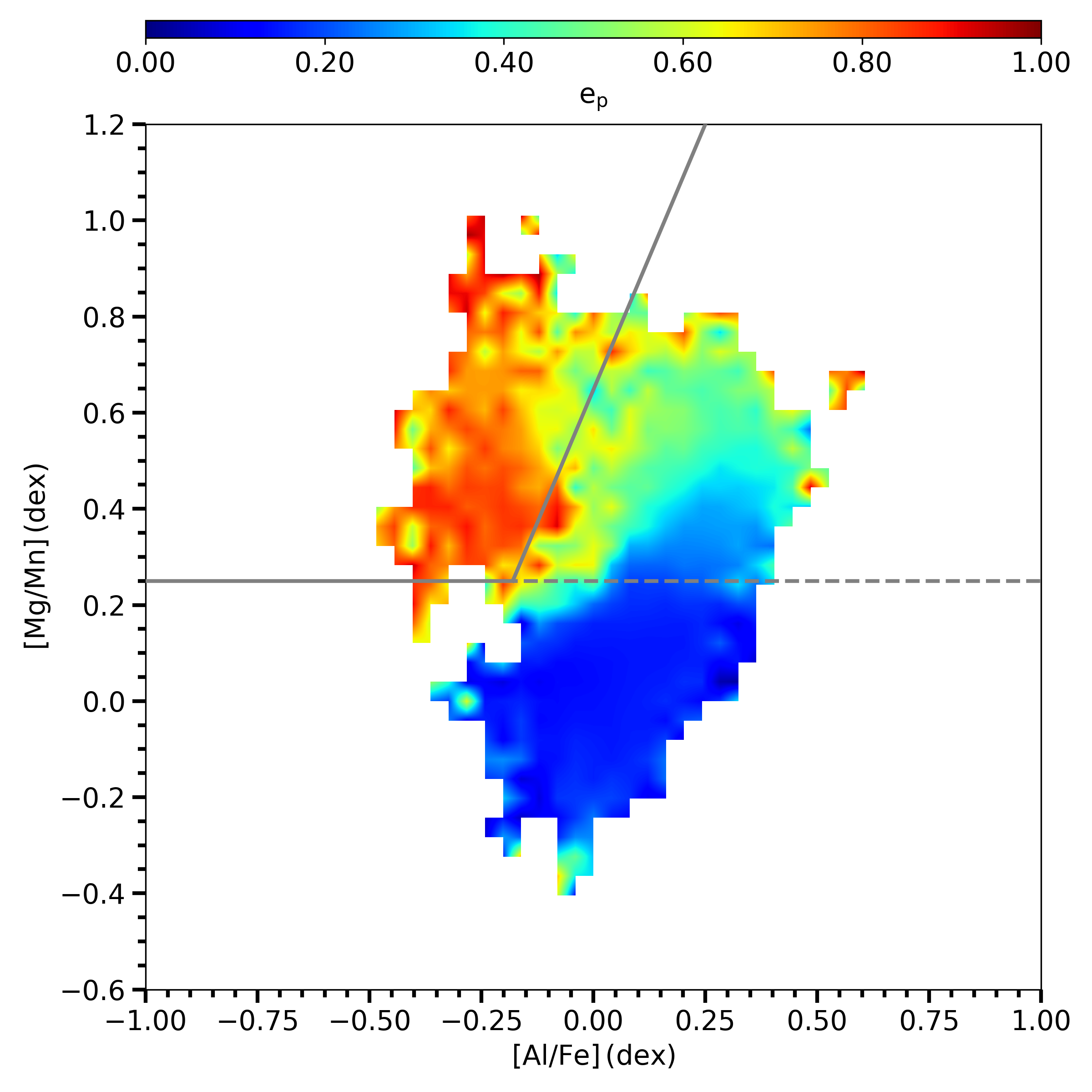}{0.30\textwidth}{}}
    \gridline{\fig{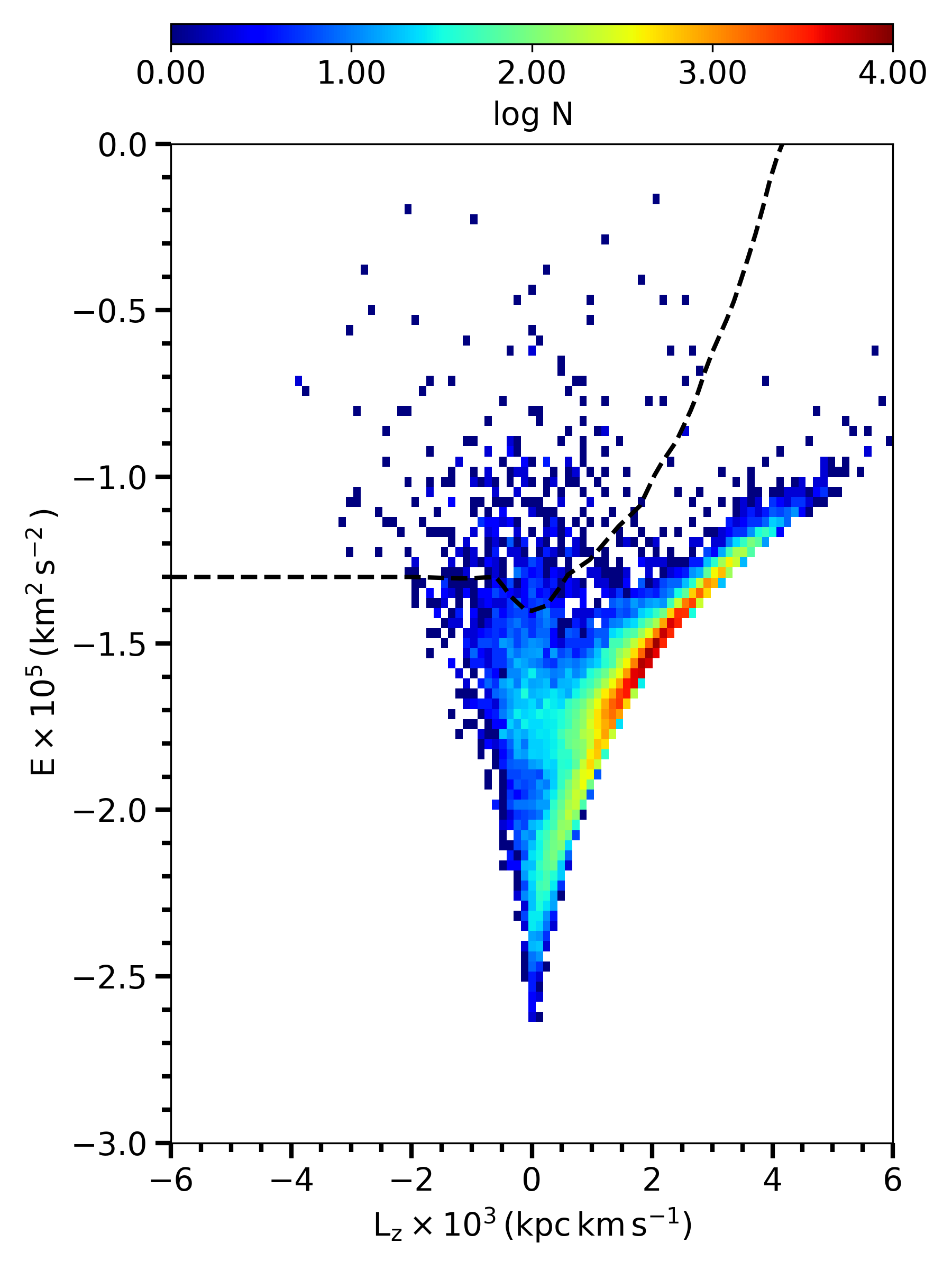}{0.32\textwidth}{}
          \fig{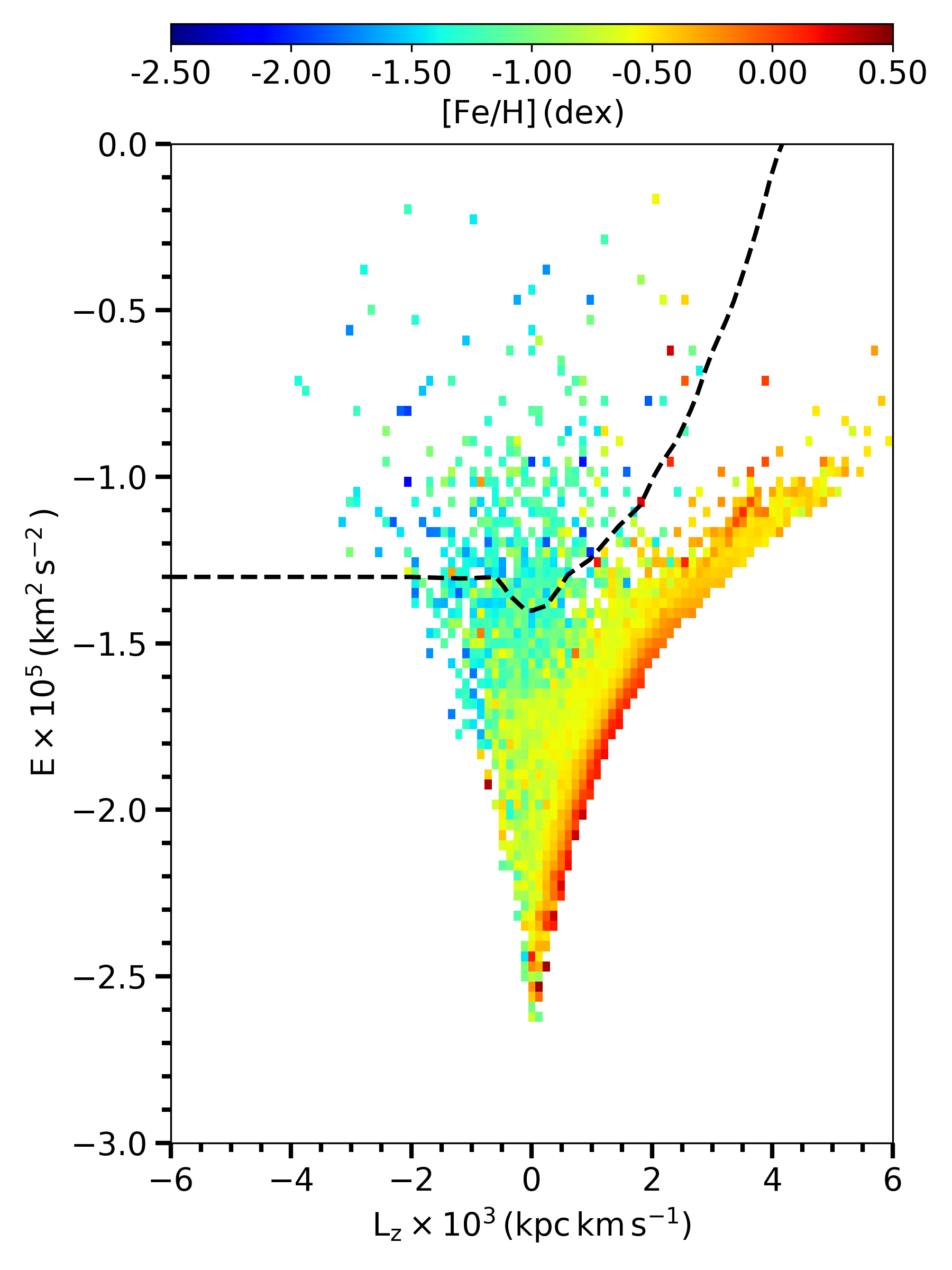}{0.32\textwidth}{}
          \fig{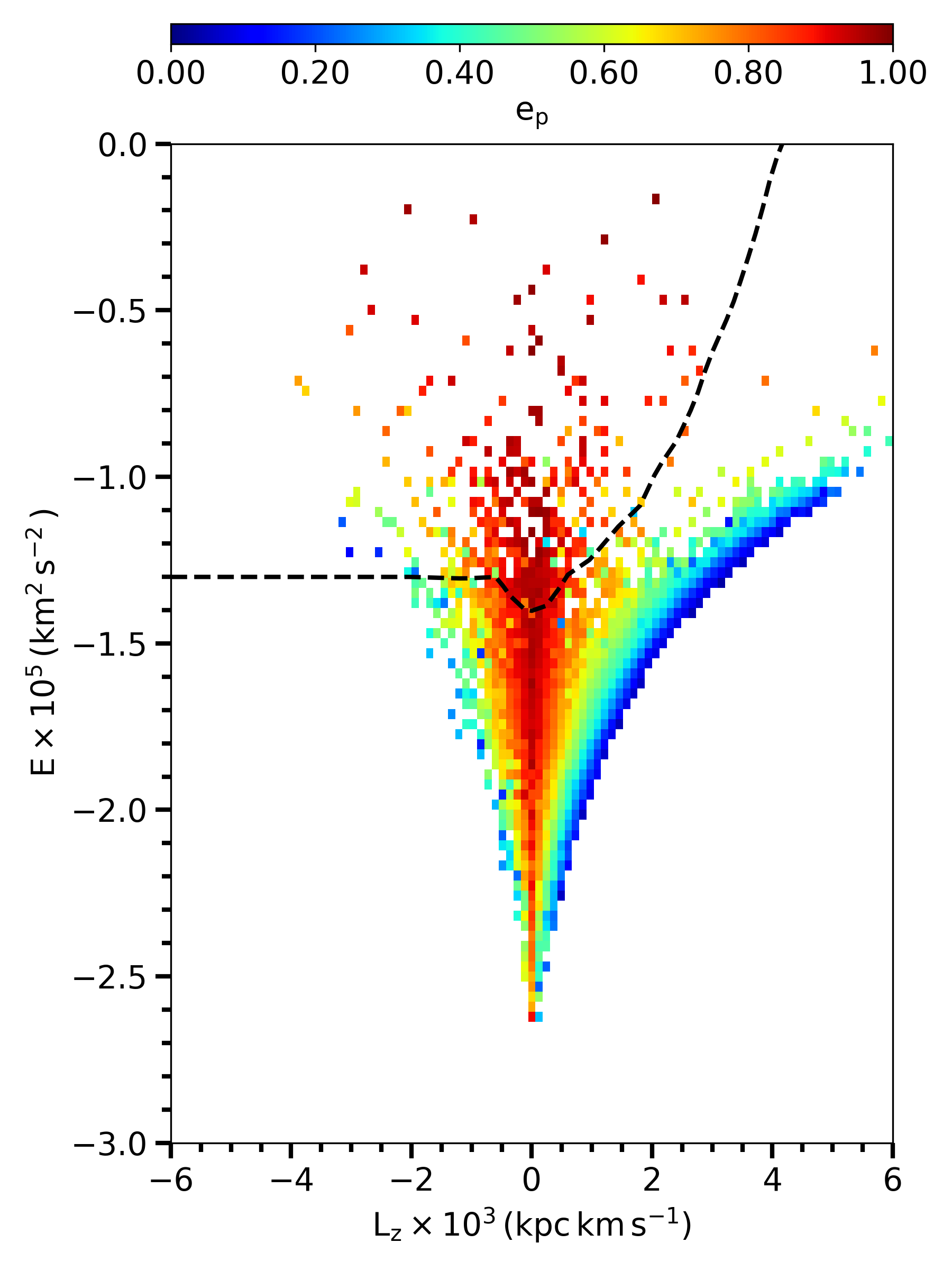}{0.32\textwidth}{}}
\caption{The Top Row: The $\mathrm{[Mg/Mn]\times[Al/Fe]}$ diagram of the selected giant stars. The solid gray and the dashed lines are taken from \citet{Horta2023}. The left panel is color-coded by number density, the middle panel by iron abundance, and the right panel by eccentricity. The Bottom Row: Lindblad diagram obtained using the calculated orbital parameters of the selected giant stars. The left panel is color-coded by number density, the middle panel by iron abundance, and the right panel by eccentricity. The black dashed line is taken from \citet{Belokurov2024}}
    \label{fig:input_diagrams}
\end{figure*}

\section{Data Selection} \label{sec:data_selection}

In this study, we utilized data from the APOGEE sky survey project (APOGEE DR17; \citealp{ApogeeDR17}) conducted under the Sloan Digital Sky Survey (SDSS) and matched it with Gaia DR3 data (\citealp{GaiaDR3}). The combined dataset provides spectroscopic, astrometric, and photometric information for 733,901 stars. To ensure reliable results, we applied several quality restrictions based on the APOGEE consortium's recommendations and the specific needs of this study. Specifically, we required that the \texttt{ASPCAPFLAG}, \texttt{STARFLAG}, and \texttt{EXTRARG} values be equal to 0, and similarly, the flags for Fe, Mg, Mn, and Al were also set to 0. In addition, we imposed a minimum signal-to-noise ratio (S/N) of 70 and a relative parallax uncertainty ($\sigma_\varpi/\varpi \leq 0.2$). For the selection of giant stars, we further restricted the sample to those with surface gravities in the range $0.5 \leq \log g \leq 3.6$ and effective temperatures between 3500~K and 5500~K. These selection criteria are well accepted cuts in the literature \citep{Carillo2023, Horta2023}. We also removed all stars identified as globular cluster members from the APOGEE globular cluster catalogue \citep{apogee_gc}. These combined quality cuts resulted in a final sample of 137,046 giant stars. 

\subsection{Space Velocity and Orbital Calculations}

The space velocities, positions, proper motions, and radial velocities of the selected giant stars were converted into a galactocentric Cartesian coordinate system using the \texttt{Astropy} library \citep{Astropy2013, Astropy2018}. For this calculation, we adopted a solar Galactocentric distance of $R_{\odot} = 8.2$ kpc \citep{BlandHawthorn2016, Gravity2019} and a local standard of rest (LSR) velocity of $V_{\mathrm{LSR}} = 232.8$ km s$^{-1}$. Additionally, the peculiar motion of the Sun with respect to the LSR was taken as $(U, V, W) = (11.10, 12.24, 7.25)$ km s$^{-1}$ \citep{Schonrich2010}. 

The Galactic orbits of the stars were then computed by integrating their equations of motion under the McMillan potential \citep{McMillan2017} using the \texttt{galpy} \citep{galpy} python library. As a result, orbital parameters for each star — including apocentric distance ($R_{\rm apo}$), pericentric distance ($R_{\rm peri}$), orbital eccentricity ($e_{\rm p}$), maximum vertical excursion ($Z_{\rm max}$), angular momentum in three dimensions ($L_{\rm x}, L_{\rm y}, L_{\rm z}$), action angles ($J_R, J_\phi, J_z$), orbital energy ($E$) and orbital energy under the circular orbit assumption ($E_{\rm circular}$) — were determined. Lindblad diagrams ($E \times L_z$) constructed from these orbital parameters are shown in the bottom row of Figure~\ref{fig:input_diagrams}, where the plots are color-coded from left to right by number density, $\mathrm{[Fe/H]}$ abundance, and orbital eccentricity, respectively.

Figure~\ref{fig:input_diagrams} illustrates the distribution of our selected dataset across key chemical and dynamical spaces, highlighting its mixed in-situ and ex-situ stellar populations. The top row presents the $ \mathrm{[Mg/Mn] \times [Al/Fe]}$ plane, a well-established chemical diagnostic used to distinguish accreted stars from those formed in situ within the Milky Way \citep{Horta2023}. The first panel represents the number density of stars in this space, showing a concentration of chemically homogeneous stars with a distinct extension toward a more metal-poor and likely ex-situ population. The second panel displays the variation in $\mathrm{[Fe/H]}$, revealing a clear metallicity gradient that suggests the presence of both metal-rich in-situ stars and metal-poor accreted stars. The third panel highlights the distribution of orbital eccentricity, with higher-eccentricity stars preferentially occupying the region associated with accreted populations.

The bottom row shows the Lindblad diagram, where energy and angular momentum serve as key dynamical tracers of accretion history. The first panel represents the number density of stars, outlining the characteristic chevron-like shape of the Milky Way’s stellar halo, with a significant presence of retrograde and radially dominated orbits. The second panel illustrates the variation in $\mathrm{[Fe/H]}$, further confirming the coexistence of metal-rich in-situ stars and metal-poor accreted components. The third panel displays the eccentricity distribution, showing that the majority of highly eccentric stars are dynamically distinct from the in-situ population. Figure~\ref{fig:input_diagrams} shows that our sample contains both in-situ and ex-situ populations, both chemically \citep{Horta2023} and kinematically \citep{Belokurov2024}. The presence of a dynamically hot, radially biased population strongly suggests that our dataset contains merger debris, making it a promising sample for identifying Gaia–Enceladus stars in the subsequent analysis.

\section{Selecting Gaia-Enceladus Stars with Machine Learning Methods} \label{sec:ml_pipeline}

Previous studies have shown that the Milky Way has accreted not only the Gaia–Enceladus dwarf galaxy, but also several other extragalactic systems \citep[e.g.,][]{Helmi2018,Horta2023,Carrillo2024}. By focusing on the kinematic and chemical properties of stars originating outside our Galaxy, these studies have identified the points at which such stars diverged from the Milky Way using a variety of analytical techniques.
In doing so, they have successfully distinguished extragalactic stars and revealed evidence of multiple merger events.

\citet{Carrillo2024} demonstrated that selecting Gaia–Enceladus stars using traditional constraints—whether based on kinematics, orbital parameters, or chemical abundances—often results in significant sample mixing. Due to overlapping features in the multidimensional parameter space, conventional selection criteria tend to confuse accreted Gaia-Enceladus stars with the in-situ stellar population. This blending can introduce systematic biases in the inferred properties of the progenitor galaxy, such as its stellar mass and chemical evolution history. To address these challenges, we adopt an unsupervised machine learning approach to improve the selection of Gaia-Enceladus stars. By integrating spectroscopic, kinematic, and chemical data within an advanced classification framework, our method significantly reduces sample mixing and increases population homogeneity. This approach stands out as an effective and reliable tool for distinguishing stellar populations that exhibit overlapping features in complex, multidimensional spaces.

\begin{figure}
    \centering
    \includegraphics[width=\columnwidth]{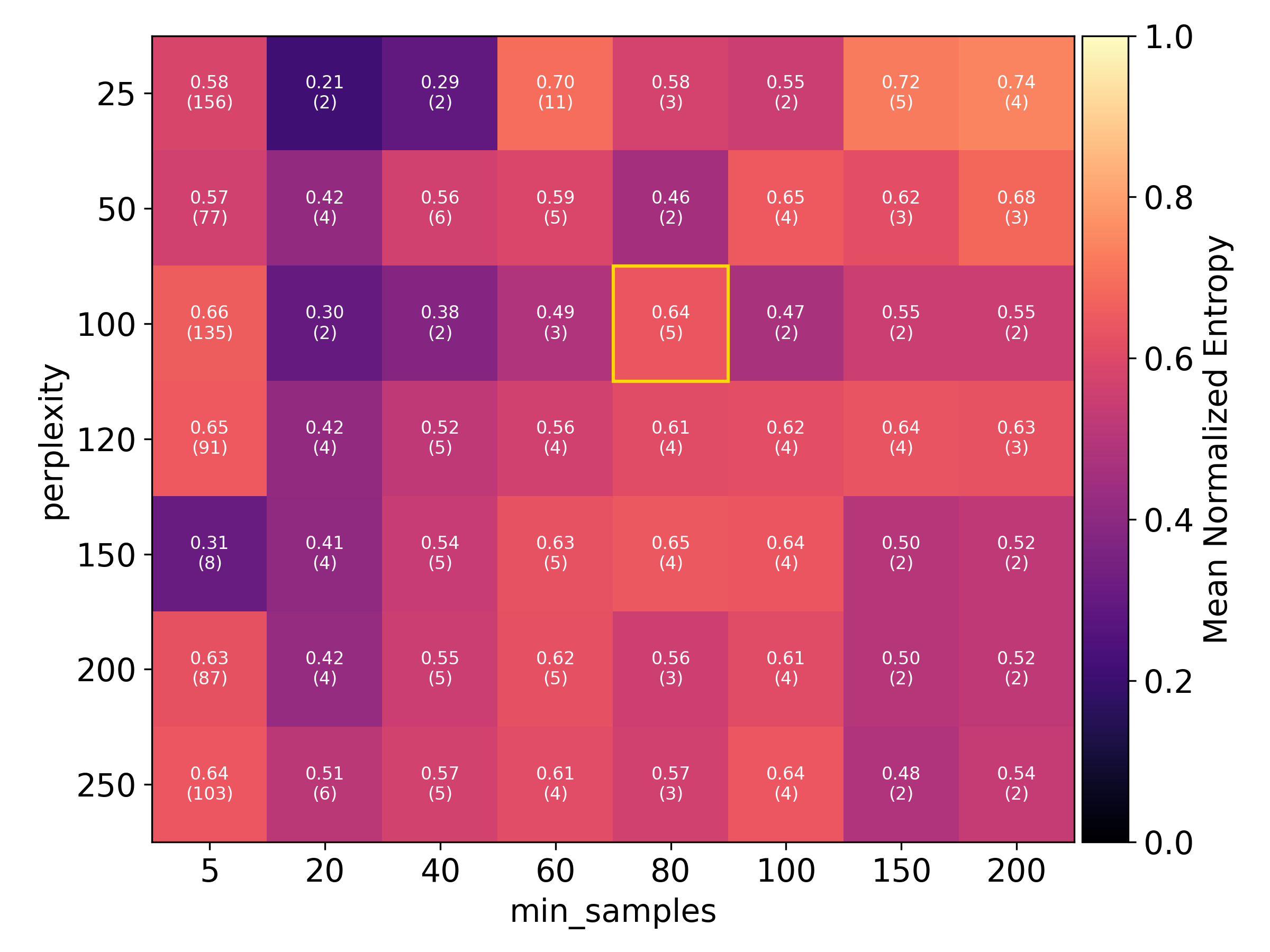}
    \caption{Mean normalized entropy values for different combinations of t-SNE perplexity and HDBSCAN \textit{min\_samples} parameters. Each grid cell reports the entropy value along with the number of clusters detected in parentheses. The yellow-bordered cell highlights the selected configuration.}
    \label{fig:normalized_entropy}
\end{figure}

\begin{table*}
\centering
\caption{Orbital and chemical properties of clusters.}
\label{tab:cluster_orbital_chem_summary}
\begin{tabular}{lcccccccccccc}
\hline
Cluster & $N_\mathrm{stars}$ & $R_\mathrm{apo}$ & $R_\mathrm{peri}$ & $R_\mathrm{mean}$ & $e$ & $Z_\mathrm{max}$ & $L_z$ & $E$ & $\mathrm{[Fe/H]}$ & $\mathrm{[Mg/Fe]}$ & $\mathrm{[Mn/Fe]}$ & $\mathrm{[Al/Fe]}$ \\
        &                     & \small{(kpc)}    & \small{(kpc)}     & \small{(kpc)}     &     & \small{(kpc)}    & \small{($10^3 \times$ kpc km s$^{-1}$)} & \small{($10^5 \times$ km$^2$ s$^{-2}$)} & \small{(dex)} & \small{(dex)} & \small{(dex)} & \small{(dex)} \\
\hline
1 & 893 & 3.79 & 0.54 & 2.18 & 0.73 & 2.05 & 0.10 & -2.16 & -0.53 & 0.33 & -0.11 & 0.21 \\
2 & 18 & 4.28 & 1.05 & 2.72 & 0.66 & 3.22 & -0.18 & -2.05 & -1.39 & 0.29 & -0.31 & -0.09 \\
3 & 1814 & 9.42 & 1.45 & 5.40 & 0.73 & 3.73 & 0.45 & -1.72 & -0.58 & 0.34 & -0.14 & 0.26 \\
4 & 28 & 10.20 & 0.61 & 5.51 & 0.88 & 5.45 & -0.14 & -1.69 & -0.70 & 0.15 & -0.16 & 0.00 \\
5 & 1124 & 14.72 & 1.13 & 8.14 & 0.87 & 8.84 & -0.14 & -1.50 & -1.27 & 0.19 & -0.32 & -0.23 \\
\hline
\end{tabular}
\label{tab:cluster_orbital_summary}
\end{table*}

The distribution of the data in the $\mathrm{[Mg/Mn]\times[Al/Fe]}$ plane and in the Lindblad diagram is shown in the top and bottom rows of Figure~\ref{fig:input_diagrams}, respectively. A close inspection of these figures reveals that stars associated with the Galactic disk overwhelmingly dominate the dataset. During our iterative analysis, we found that the density-based clustering algorithm (HDBSCAN) we employed was strongly influenced by these dominant disk features, which hindered our ability to reliably detect the relatively scarce Gaia–Enceladus stars and other merger signatures. Therefore, before attempting to isolate Gaia-Enceladus stars in the data, we examined the $e_p$ color-coding presented in the third column of Figure~\ref{fig:input_diagrams} and inferred that imposing a condition of $e_p\leq0.6$ could effectively segregate the disk population from others. As a result, we removed data from the dataset that satisfied the constraints $L_z \geq 0$ and $e_p\leq0.6$, reducing our sample to 4,987 stars.

For the machine learning pipeline used to isolate Gaia–Enceladus stars from our dataset, we employed six key parameters—$\mathrm{[Fe/H]}$, $\mathrm{[Mg/Fe]}$, $\mathrm{[Al/Fe]}$, $\mathrm{[Mn/Fe]}$, orbital energy ($E$), and the angular momentum component along the Galactic $z$-axis ($L_z$)—that combined two crucial discriminating planes, as illustrated in Figure~\ref{fig:input_diagrams}. 

In our method, we first reduced the six-dimensional parameter space to a two-dimensional projection using the t-SNE algorithm \citep{tsne}. This algorithm performs dimensionality reduction by constructing a t-distribution based on the pairwise similarities between data points in high-dimensional space, thereby preserving local and global structure more effectively than linear techniques such as PCA. A key hyperparameter of t-SNE is the perplexity, which controls the balance between sensitivity to local versus global structure by determining the number of effective neighbors considered in the embedding. Small perplexity values highlight fine-grained local structure and can result in over-fragmentation, whereas large values tend to preserve global trends but risk blurring compact substructures. As t-SNE is a stochastic, unsupervised projection technique, selecting an optimal perplexity value is non-trivial and strongly dependent on both the dataset’s intrinsic structure and the clustering algorithm applied downstream.

\begin{figure}
    \centering
    \includegraphics[width=\columnwidth]{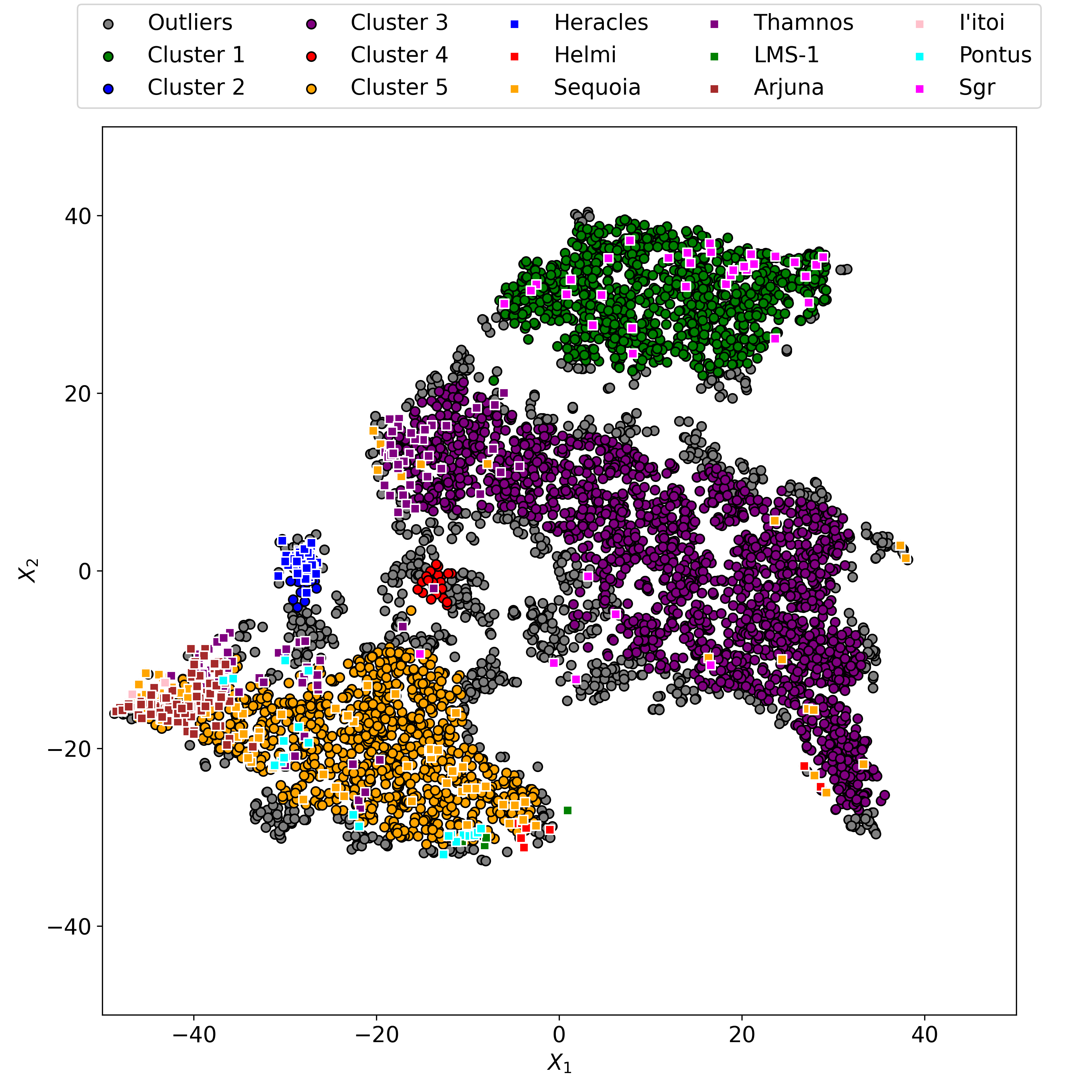}
    \caption{Two-dimensional projection of the stellar sample obtained via t-SNE using six chemical abundance ratios as input. Colored circles indicate the five clusters and outliers identified by the HDBSCAN algorithm. Overlaid colored squares represent stars associated with previously known Galactic substructures, based on chemo-dynamical criteria from the literature.}
    \label{fig:tsne_hdbscan}
\end{figure}

To objectively determine a suitable combination of t-SNE and HDBSCAN parameters, we evaluated clustering outcomes using the mean normalized entropy, an internal validation metric that quantifies the dispersion of cluster member properties in physically meaningful spaces. Lower entropy indicates greater cluster coherence and less overlap across features such as chemical abundances and orbital parameters, implying that the algorithm successfully identifies astrophysically interpretable substructures. Conversely, higher entropy values suggest poorer separation, with substantial mixing across clusters. This is especially critical in unsupervised workflows where no external ground truth exists.

We systematically explored all parameter combinations yielding mean normalized entropy values between 0.30 and 0.70, covering a broad range of perplexity and HDBSCAN’s \textit{min\_samples} values. Figure~\ref{fig:normalized_entropy} displays the entropy landscape for these configurations, with each grid cell indicating the resulting mean entropy and the corresponding number of clusters. While several parameter sets yielded slightly lower entropy values (e.g., perplexity = 50 or 120, min\_samples = 100), we found through visual inspection that these configurations often produced fewer clusters or merged chemically distinct populations. This highlights that entropy minimization alone does not guarantee astrophysically meaningful separation, and additional diagnostics are required.

To move beyond entropy minimization and assess the astrophysical significance of the clustering, we explicitly generated and inspected diagnostic visualizations for every individual parameter combination. Figure~\ref{fig:tsne_hdbscan} illustrates the resulting t-SNE projection with color-coded cluster assignments, highlighting dominant and minor substructures. We then examined the high-dimensional clustering in the input abundance space (Figure~\ref{fig:clustered_input_space}) and its projections onto key chemical and orbital planes (Figure~\ref{fig:diagnostic_graphs}). These visual inspections allowed us to evaluate how effectively each configuration separated chemically and dynamically coherent stellar groups. The configuration with perplexity = 100 and \textit{min\_samples} = 80, yielding a moderate entropy value of 0.64, was found to best isolate the Gaia–Enceladus population with minimal overlap and maximal contrast relative to other stellar structures. This setting produced five chemically and dynamically interpretable clusters, including a group consistent with the expected properties of accreted stars. The median chemical and dynamical properties, as well as the number of stars in each resulting cluster, are summarized in Table~\ref{tab:cluster_orbital_chem_summary}.

\section{Identification Gaia-Enceladus Stars inside from HDBSCAN clusters} \label{sec:humanization}

\begin{figure*}
    \centering
    \gridline{\fig{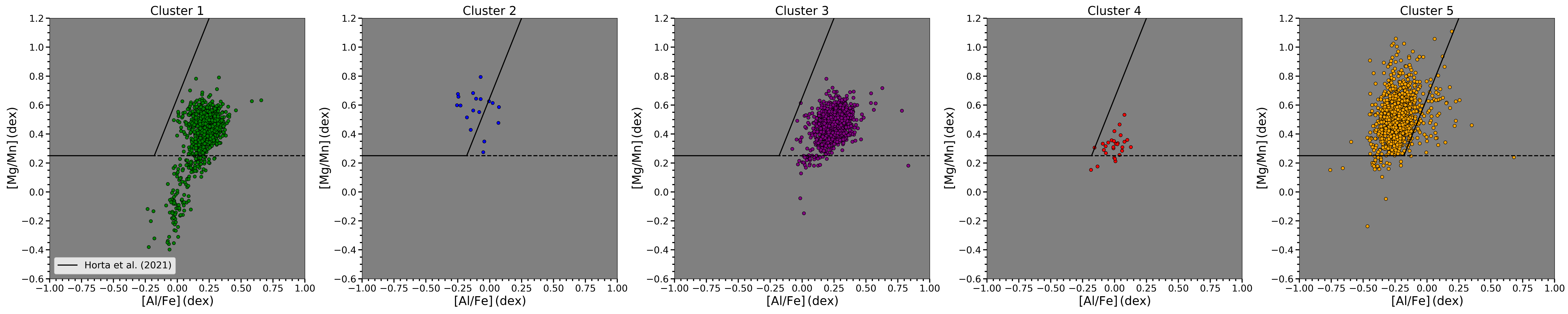}{\textwidth}{}}\vspace{-1cm}
    \gridline{\fig{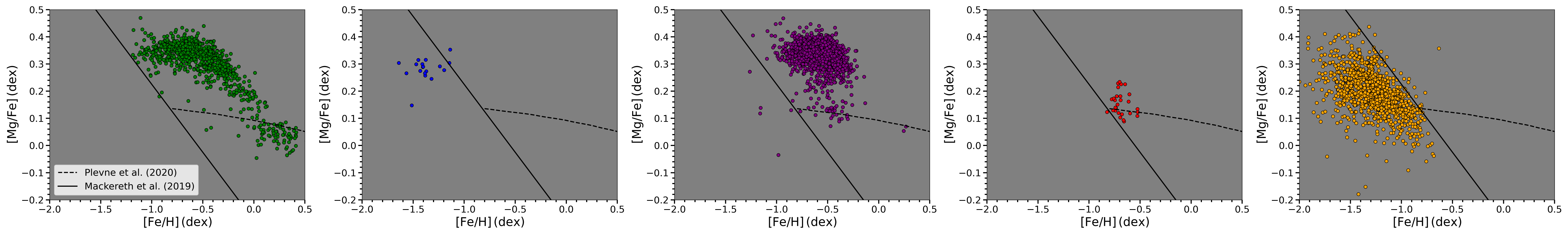}{\textwidth}{}}\vspace{-1cm}
    \gridline{\fig{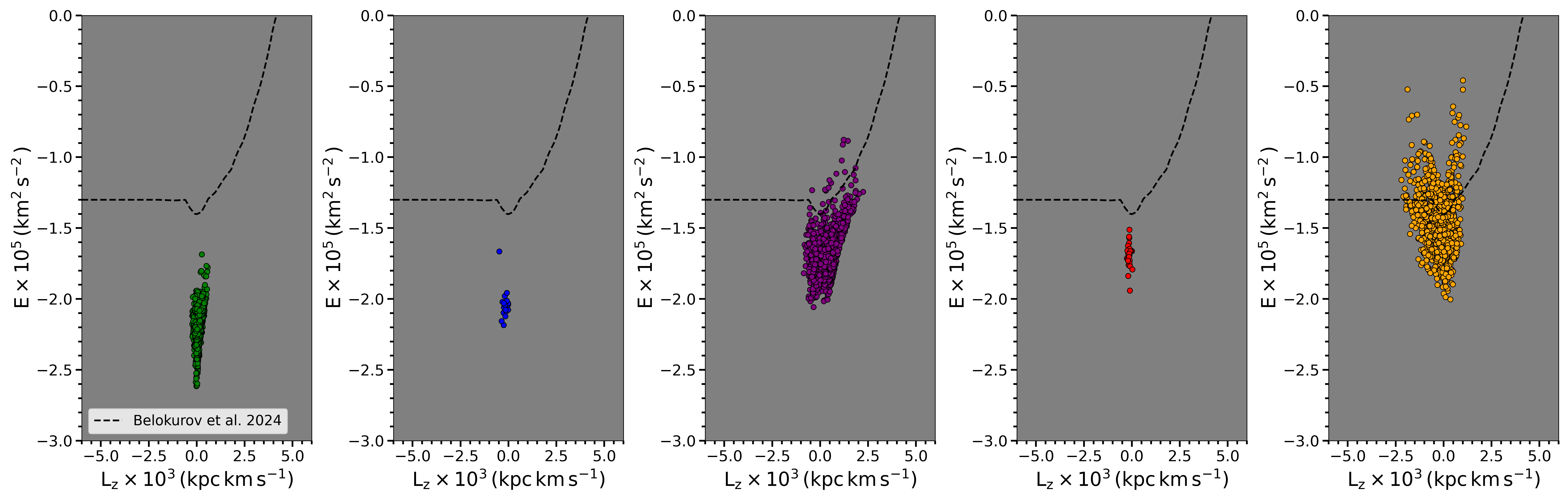}{\textwidth}{}}\vspace{-0.75cm}
    \caption{The distributions of the clustering results in the $\mathrm{[Mg/Mn]\times[Al/Fe]}$ chemical plane (upper panel), $\mathrm{[Mg/Fe]\times[Fe/H]}$ chemical plane (middle panel) and their distributions in the $E \times L_z $ kinematic plane (lower panel).}
    \label{fig:clustered_input_space}
\end{figure*}

Since HDBSCAN is an unsupervised machine learning algorithm that hierarchically groups data based on their distribution, it is essential to determine which Galactic populations or merger remnants the resulting clusters represent. To assign population membership, we examined the distributions of the input parameters across different projection planes, as illustrated in Figure~\ref{fig:clustered_input_space}. This figure demonstrates that the five clusters identified by HDBSCAN in the t-SNE projected space are effectively segregated into distinct groups. At first, these clusters appear to discriminate between in-situ and ex-situ stellar populations: specifically, Clusters 1 and 3 predominantly consist of stars native to the Milky Way, Cluster 5 exhibits the distribution expected for Gaia–Enceladus stars, and Clusters 2 and 4 likely correspond to stellar groups originating from other merger events. To further investigate these associations and explore possible correspondences with known Galactic substructures, we systematically examined the chemical and dynamical properties of the clustered dataset using literature-based criteria.

\begin{figure*}
    \centering
    \includegraphics[width=\textwidth]{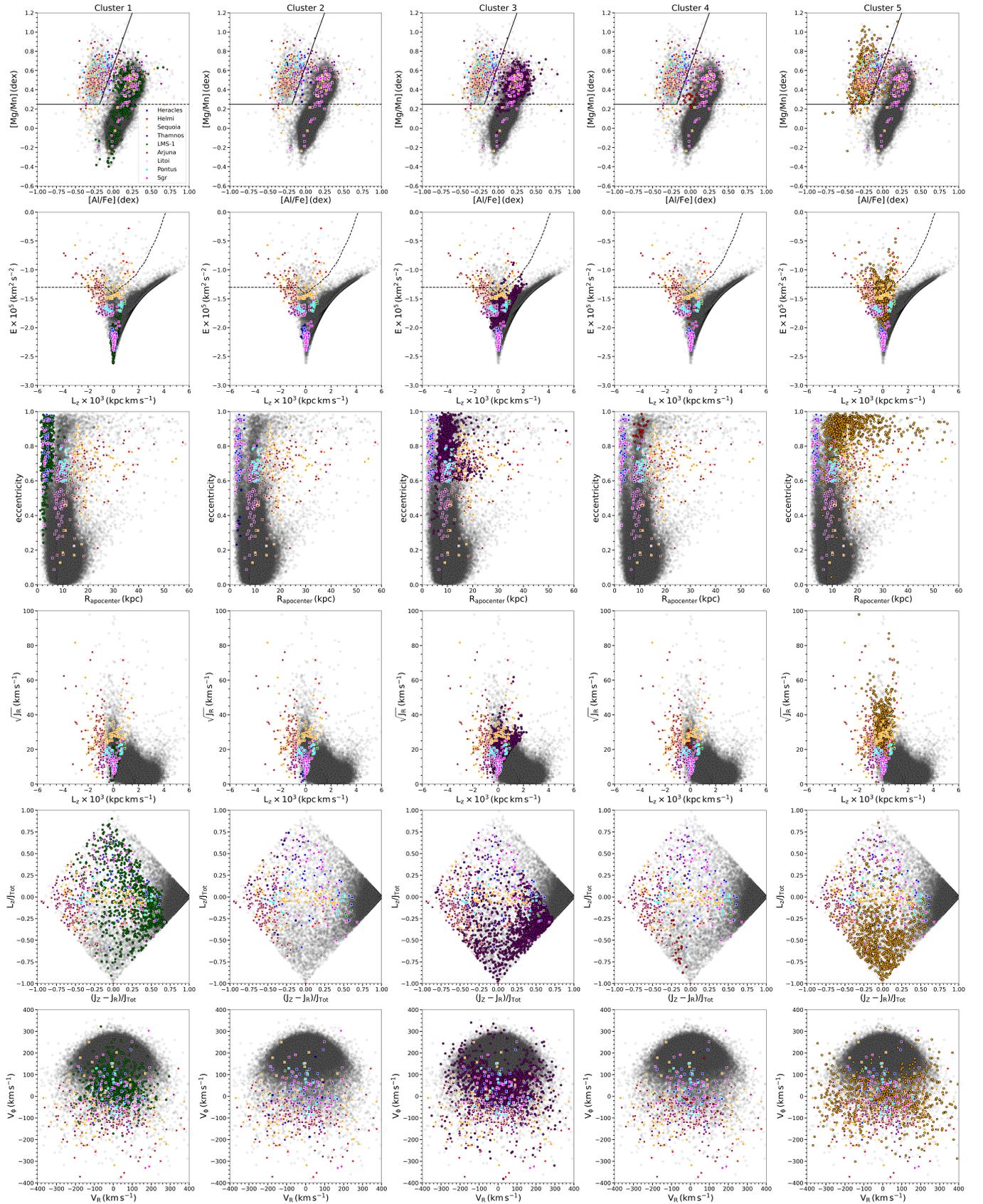}
    \caption{The clusters identified by the HDBSCAN algorithm are color-coded and overlaid on the clustring dataset (shown in grayscale) to highlight their separation in different feature spaces. Additionally, known accreted populations such as Heracles, Helmi, Sequoia, Thamnos, LMS, Arjuna, I’itoi, Pontus and Sagittarius are identified with unique square markers.}
    \label{fig:diagnostic_graphs}
\end{figure*}

However, a more robust determination of the population membership of these clusters requires evaluating them in additional parameter spaces and comparing the results with the constraints for various merger remnants provided in the literature. For this reason, we applied criteria for several accretion events outlined by \citep{Horta2023} to the entire dataset used for clustering. The outcomes of these diagnostic tests, compiled across several projection planes, are presented in Figure~\ref{fig:diagnostic_graphs}. Each column represents a distinct projection of the dataset, where the individual panels display the relationship between various kinematic and chemical parameters. The clusters identified by the HDBSCAN algorithm are color-coded and overlaid on the clustering dataset and disc population (shown in grayscale) to highlight their separation in different feature spaces. Additionally, known accreted populations such as Heracles \citep{Horta2021}, Helmi Streams \citep{Helmi_Streams}, Sequoia \citep{Barba2019, Matsuno2019, Myeong2019}, Thamnos \citep{Koppelman2019}, LMS \citep{Yuan2020}, Pontus \citep{Malhan2022}, Saggitarius \citep{Ibata1994}, Arjuna and I’itoi \citep{Naidu2020}, are identified with unique square markers. These visualizations provide insight into the dynamical and chemical characteristics of each group, aiding in the interpretation of merger remnants and their distinction from the Milky Way’s in-situ populations.

Based on our clustering analysis, the final sample includes stars associated with several distinct Galactic populations. It is important to note that the numbers reported below do not correspond to individual HDBSCAN clusters, but rather to identifications made within our sample using the chemo-dynamical selection criteria compiled in the literature summary tables of \citet{Horta2023}. Specifically, we identified 44 stars belonging to the Heracles population, 13 to Helmi, 126 to Sequoia, 107 to Thamnos, 7 to LMS, 109 to Arjuna, 2 to I’itoi, 27 to Nyx, 21 to Pontus, and 37 to Sagittarius.

As shown in Figure~\ref{fig:tsne_hdbscan}, several literature-defined substructures are visibly concentrated in projection space but were not separated as individual clusters by the algorithm. This reflects the limitations of the chosen hyperparameters in detecting numerically small but chemically distinct groups. For example, although the Heracles population appears compact in the t-SNE projection, it was absorbed into Cluster 2, suggesting a connection between the two. In contrast, many of the remaining substructures (Thamnos, Sequoia, Arjuna, LMS-1, Helmi) are distributed within Cluster 5. This overlap reflects the intrinsic challenge of isolating substructures with overlapping abundance-orbit distributions in unsupervised clustering frameworks.

Given the primary goal of this study, to isolate Gaia–Enceladus stars, we prioritized configurations that yielded a dominant cluster with properties aligned with the expected chemical and dynamical signatures of this merger remnant. To further refine the selection and eliminate contamination from overlapping substructures, we cross-matched our clustered sample with known literature-defined components, ultimately isolating a chemically clean subset of Gaia-Enceladus candidates.

\begin{figure}
    \centering
    \includegraphics[width=\columnwidth]{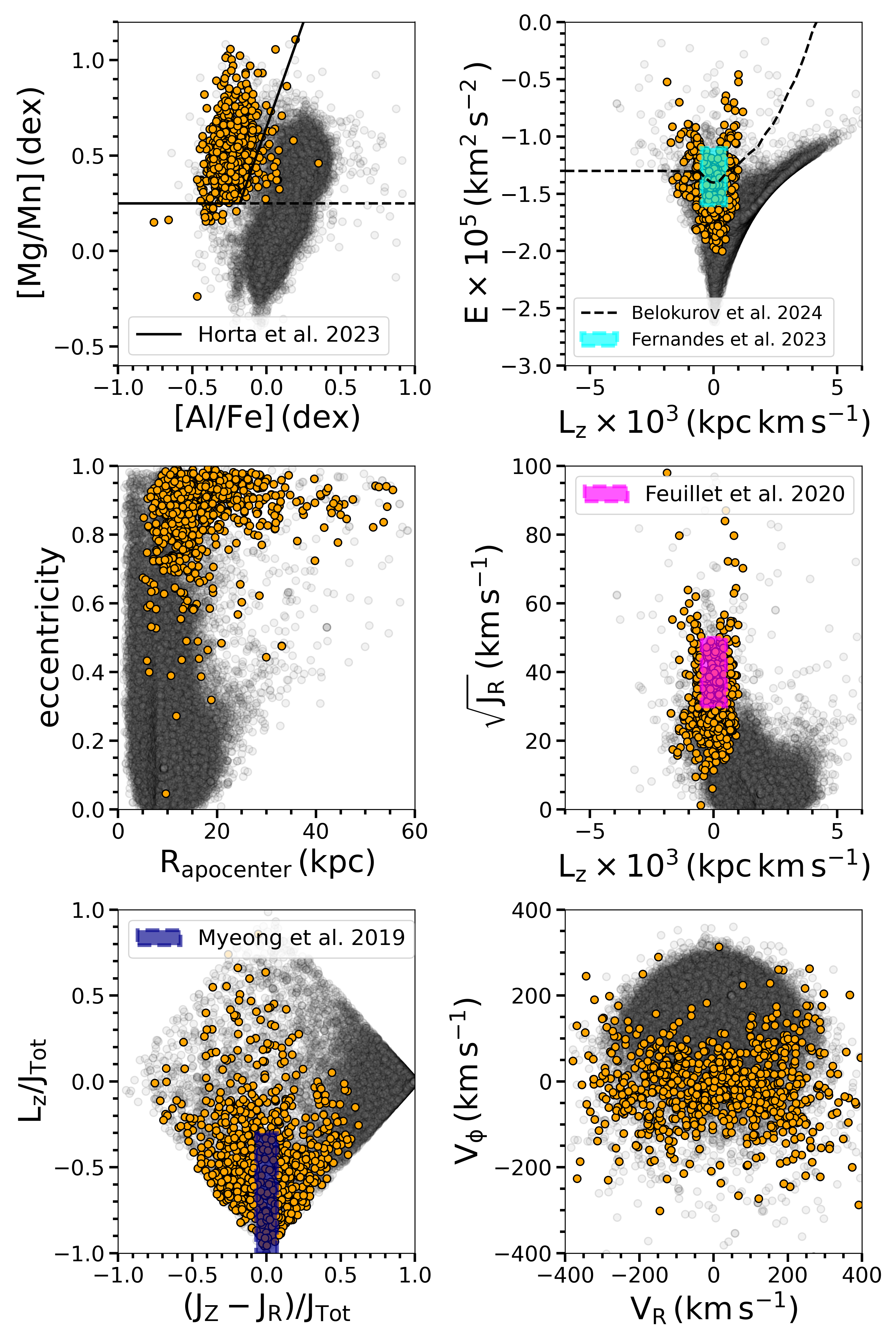}
    \caption{Diagnostic diagrams illustrating the distribution of the Gaia–Enceladus candidate stars selected in this study with orange points across various chemical and dynamical planes. The background gray points corresponds to the full sample of the study. Colored rectangular regions represent Gaia–Enceladus selection boxes proposed in the literature for each respective plane.}
    \label{fig:GES_diagnostic}
\end{figure}

We examined the results of the machine learning pipeline employed to isolate Gaia–Enceladus stars from our dataset by comparing them with literature-based criteria across various parameter planes. Our analysis shows that the stars grouped into Cluster 5 (orange) exhibit distributions consistent with Gaia–Enceladus. After excluding those Cluster 5 stars allocated to other literature-identified populations, the machine learning pipeline indicates that 884 stars joined the Milky Way through the Gaia–Enceladus merger event. Figure~\ref{fig:GES_diagnostic} illustrates the distribution of these stars in the same planes previously shown in Figure~\ref{fig:diagnostic_graphs}.

As described above, the clustering analysis suggests that the majority of stars in Cluster 5 are likely associated with the Gaia-Enceladus population, which is the primary focus of this study. However, to substantiate this hypothesis, it is essential to examine Figure~\ref{fig:GES_diagnostic}. The first row of this figure displays the principal chemical and dynamical dimensions that underlie the integrated clustering space. The top-left panel reveals that Cluster 5 stars predominantly occupy the ex-situ region, a distinguishing feature of Gaia-Enceladus stars that has been supported by several previous studies \citep{Hawkins2015, Das2020, Horta2020, Fernandes2023}. In the top-right panel, the Lindblad diagram shows that Cluster 5 stars are widely distributed across a broad energy range at very low angular momentum. These stars span both sides of the in-situ and ex-situ dynamical separation proposed by \citet{Belokurov2024}, and they completely fill the selection box (shown in cyan) defined by \citet{Fernandes2023} for identifying Gaia-Enceladus stars. This distribution further supports the association of Cluster 5 with Gaia-Enceladus.

The middle-left panel reveals a distinct distribution at high eccentricities and large apocentric distances, which is consistent with the radial merger scenario proposed by \citet{Helmi2018}. The distribution in the middle-right panel also aligns with this merger scenario, and encompasses the expected region (highlighted by the purple box) suggested by \citet{Feuillet2020}, while extending beyond it.

In the bottom-left panel, the action-diamond diagram shows that Cluster 5 stars are well represented within the selection box proposed by \citet{Myeong2019}, while also spreading across a wider region—consistent with other literature results. Finally, in the bottom-right panel, the velocity-space distribution shows that the stars are predominantly spread within the $V_\phi$ range of –100 to 100 km/s, with a broader spread in $V_R$, capturing the expected kinematic signatures of Gaia-Enceladus stars\citep{Helmi2018,Belokurov2018}.

Each panel in Figure~\ref{fig:GES_diagnostic} illustrates a different projection space, all of which consistently demonstrate that Cluster 5 stars exhibit distributions compatible with Gaia-Enceladus, albeit in a broader and less sharply bounded manner than previously defined in the literature. Based on these comparisons, we conclude that the majority of stars in Cluster 5—identified through an unsupervised learning algorithm—likely originated from the Gaia-Enceladus dwarf galaxy.

A cross-comparison with literature-defined selection criteria indicates that Cluster 2 is largely consistent with the Heracles population, as proposed by \citet{Horta2021} and identified as a distinct population in \citet{Horta2023}. The majority of stars in this cluster occupy the expected regions in both chemo-dynamical and orbital parameter spaces characteristic of Heracles debris. As shown in Figure~\ref{fig:diagnostic_graphs}, these stars predominantly occupy the same regions as Heracles candidates, reinforcing the correspondence between Cluster 2 and this known accreted component.

\section{A Chemical Evolution Approach to Estimating the Mass of Gaia-Enceladus} \label{sec:chemev_model}

In this study, we employed the star formation version of the \textit{OMEGA+} code \citep{Cote2018} to model the chemical evolution of Gaia-Enceladus, with the primary goal of constraining its initial dwarf galaxy mass. Our framework captures the key processes that regulate galactic chemical evolution-star formation, gas inflows, gas outflows, and stellar enrichment-and relies on the following main equation for the time evolution of the gas reservoir:

\begin{align} \label{eq:main_equ}
M_{\mathrm{gas}}(t + \Delta t) &= M_{\mathrm{gas}}(t) + \dot{M}_\mathrm{inflow}(t) - \dot{M}_\ast(t) \\ \nonumber
&-\dot{M}_\mathrm{out}(t) + \dot{M}_\mathrm{ej}(t),
\end{align}

where $\dot{M}_{\mathrm{inflow}}(t)$ is the gas infall rate, $\dot{M}_\mathrm{out}(t)=\eta\,\dot{M}_\ast(t)$ the outflow rate  (proportional to the star formation rate), $\eta$ is the mass-loading factor, and $\dot{M}_\mathrm{ej}(t)$ the mass returned to the interstellar medium by evolving stars.

We adopt an exponential infall function of the form
\begin{equation} \label{eq:infall_equ}
\dot{M}_\mathrm{inflow}(t) = A_\mathrm{norm}\,\exp\!\Bigl(-\,\frac{t - \tau_\mathrm{peak}}{\tau_\mathrm{infall}}\Bigr),
\end{equation}, 

where $\tau_\mathrm{peak}$ is the epoch of maximum infall rate, and $\tau_\mathrm{infall}$ the characteristic timescale of the exponential decline. The normalization constant $A_\mathrm{norm}$ is determined as 

\begin{equation}
A_\mathrm{norm} =
\frac{M_\texttt{gas}}{
\tau_\mathrm{infall}
\Bigl[1 - \exp\Bigl(-\,\frac{\,t_\mathrm{end} - \tau_\mathrm{peak}}{\,\tau_\mathrm{infall}}\Bigr)\Bigr]
},
\end{equation}

ensuring that the integrated inflow between $t_\mathrm{start}=0$ and $t_\mathrm{end}$ matches the target accreted gas mass $M_\texttt{gas}$. While the normalization constant $A_\mathrm{norm}$ formally depends on $t_\mathrm{end}$, its effect is solely to compensate for the integration length: shorter $t_\mathrm{end}$ requires a larger normalization to achieve the same total inflow. For instance, we verified that choosing $t_\mathrm{end} = 4$ Gyr instead of 13 Gyr increases $A_\mathrm{norm}$ by nearly a factor of two. Nevertheless, since the total gas accreted is constrained to $M_{gas}$, this rescaling does not alter the net gas budget or the resulting chemical evolution. Therefore, extending the integration to 13 Gyr does not affect the physical interpretation but allows us to explore the future trajectory of Gaia–Enceladus on the chemical plane.

\begin{figure*}
    \includegraphics[width=\textwidth]{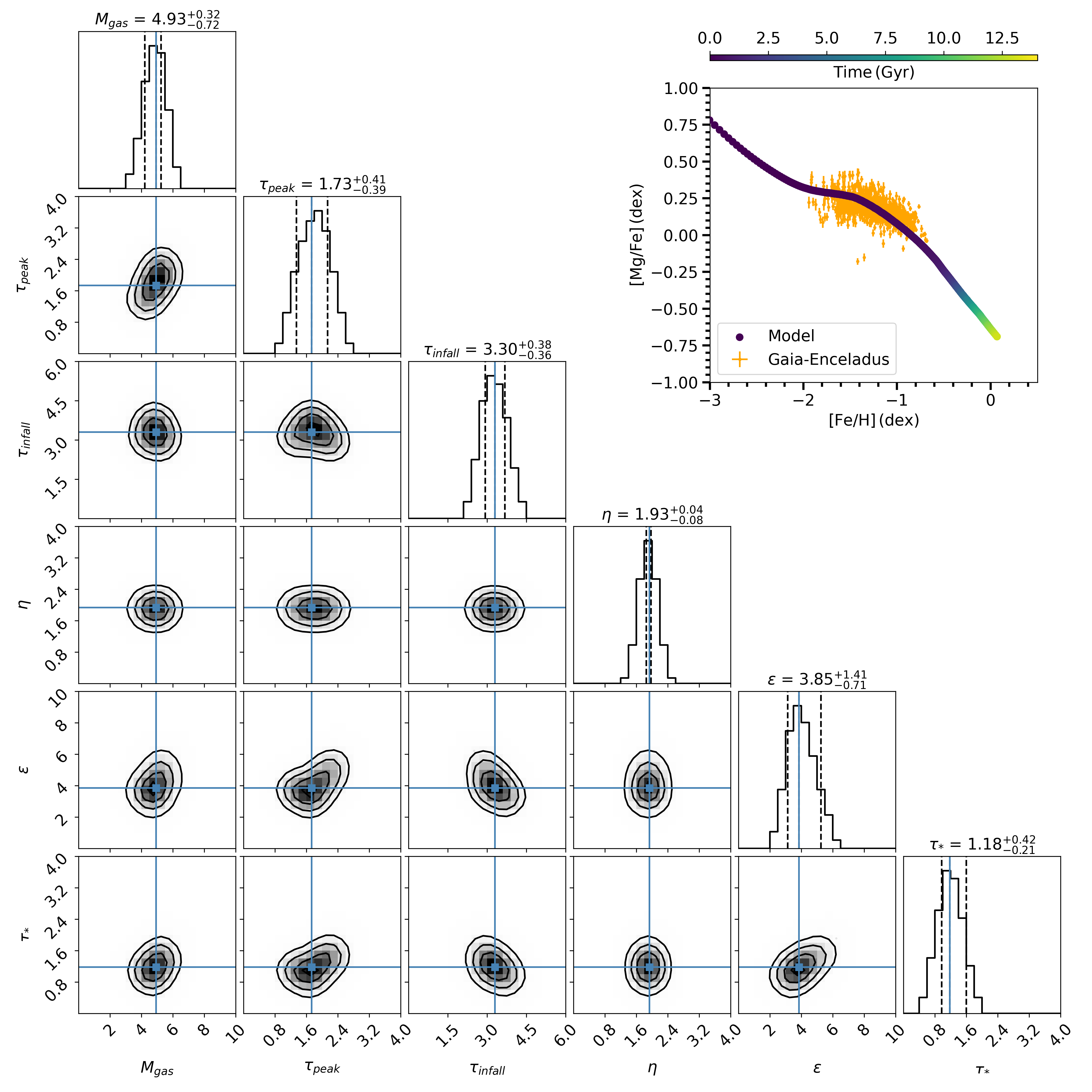}
    \caption{Posterior distributions of the chemical evolution model input parameters for Gaia-Enceladus dwarf galaxy.}\label{fig:mcmc_result}
\end{figure*}

\subsection{Star Formation and Chemical Enrichment}

The star formation rate follows the Schmidt-Kennicutt law \citep{Schmidt1959, Kennicutt1989},
\begin{equation} \label{eq:sfr_equ}
\dot{M}_\ast(t) = \frac{\epsilon \, M_{\mathrm{gas}}(t)}{\tau_\ast},
\end{equation}
where $\tau_\ast$ is the characteristic timescale to convert gas into stars, and $\epsilon$ is the star formation efficiency. Chemical enrichment is modeled using (i) the Kroupa IMF \citep{Kroupa1993}, (ii) yields from \citet{Karakas2010} for AGB stars and from \citet{Kobayashi2006} for core-collapse supernovae, and (iii) the W7 model of \citet{Iwamoto1999} plus a delay-time distribution \citep{Maoz2014} for Type Ia supernovae. 

\subsection{MCMC Fitting of Chemical Evolution Model}

To self-consistently reproduce the $\mathrm{[Mg/Fe] \times [Fe/H]}$ abundance distribution of the selected Gaia-Enceladus stars, we employed a Markov Chain Monte Carlo (MCMC) approach. The choice of this abundance plane was motivated by two main reasons. First, within the APOGEE dataset, magnesium was among the elements with the highest number of reliable measurements (i.e., flag value is 0) even before applying any element-specific quality cuts to the selected sample—alongside iron. Second, this particular plane has been widely adopted in the literature \citep{Chiappini1997, FernandezAlvar2018, Spitonietal2020, Spitonietal2021, Hasselquist2021} for both Milky Way and dwarf galaxy studies, making it a well-established and suitable basis for comparative analysis.The free parameters in our model include:

\begin{itemize}
    \item $M_{\mathrm{gas}}$: the initial gas mass in the system,
    \item $\tau_{\mathrm{infall}}$: the timescale controlling the peak of the infall function,
    \item $\tau_{\mathrm{peak}}$: the characteristic duration over which the majority of gas inflow occurs,
    \item $\tau_{\ast}$: the timescale for converting gas into stars under the Schmidt-Kennicutt law,
    \item $\epsilon$: the star formation efficiency, and
    \item $\eta$: the mass-loading factor coupling the outflow rate to the star formation rate.
\end{itemize}

We initialized 20 MCMC walkers over 5000 iterations, following a procedure similar to \citet{Spitonietal2020}. During the MCMC sampling, we employed a maximum likelihood approach to fit the model-generated abundance tracks to the observed Gaia-Enceladus stars in the $\mathrm{[Mg/Fe] \times [Fe/H]}$ plane. At each iteration, the model was integrated forward in time with the current parameter set, producing synthetic abundance distributions that were then compared against the measured values. The likelihood function quantified how closely the synthetic data aligned with observations, and parameter updates were guided by this likelihood until the Markov chains converged on an optimal solution.

All model outputs were normalized to solar abundances using the reference values of \citet{Asplund2009}. A corner diagram of the best-fitting parameter set is displayed in Figure~\ref{fig:mcmc_result}, illustrating how each parameter converges toward a credible range consistent with Gaia-Enceladus’ observed chemical and star formation properties. This final solution provides a robust lower bound on the galaxy’s pre-merger mass and underscores the power of multi-parameter chemical evolution models for interpreting ancient merger debris in the Milky Way.

\subsection{Model Predictions across Different Chemical Planes}

\begin{figure*}
    \gridline{\fig{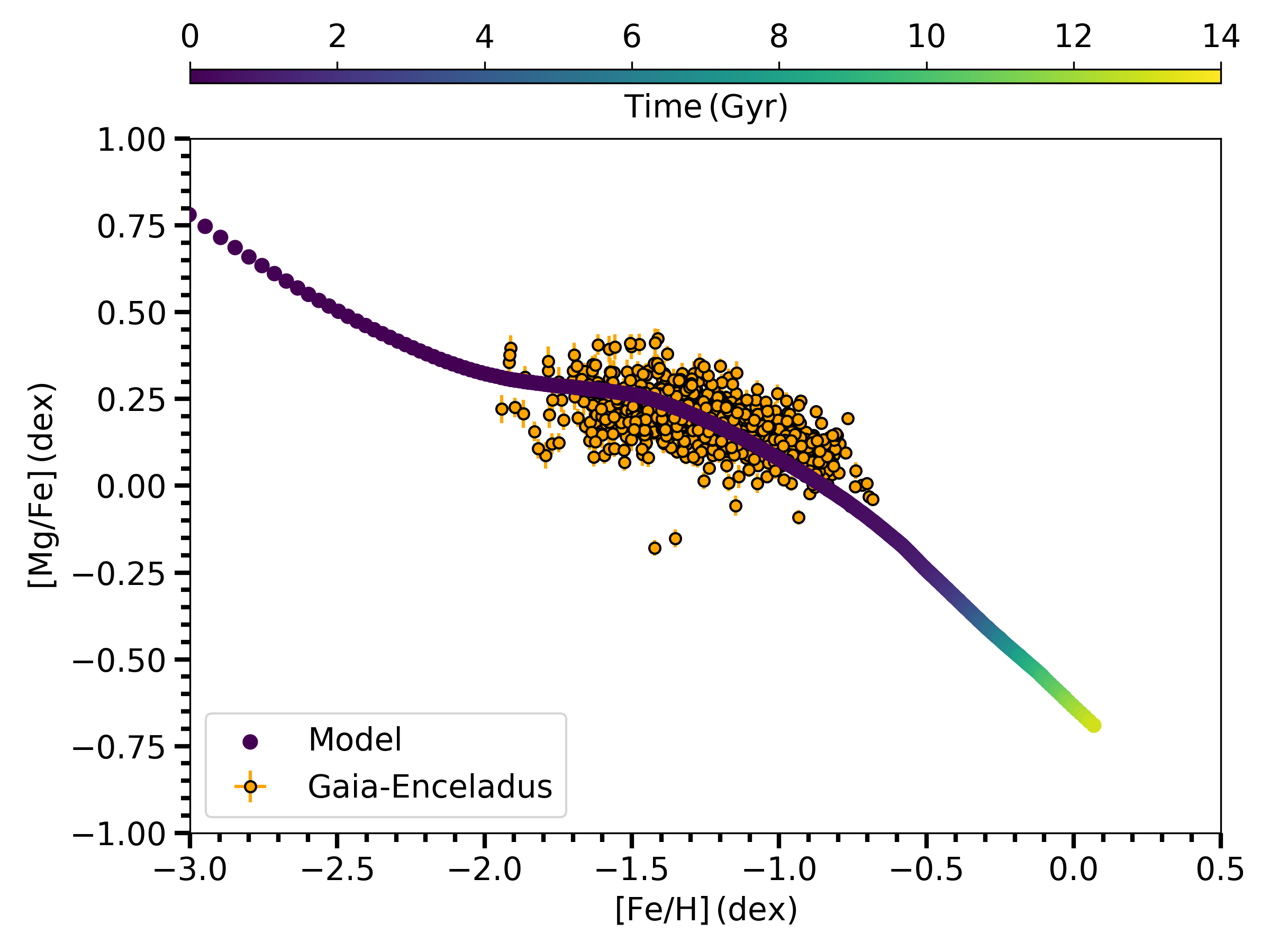}{0.30\textwidth}{}
               \fig{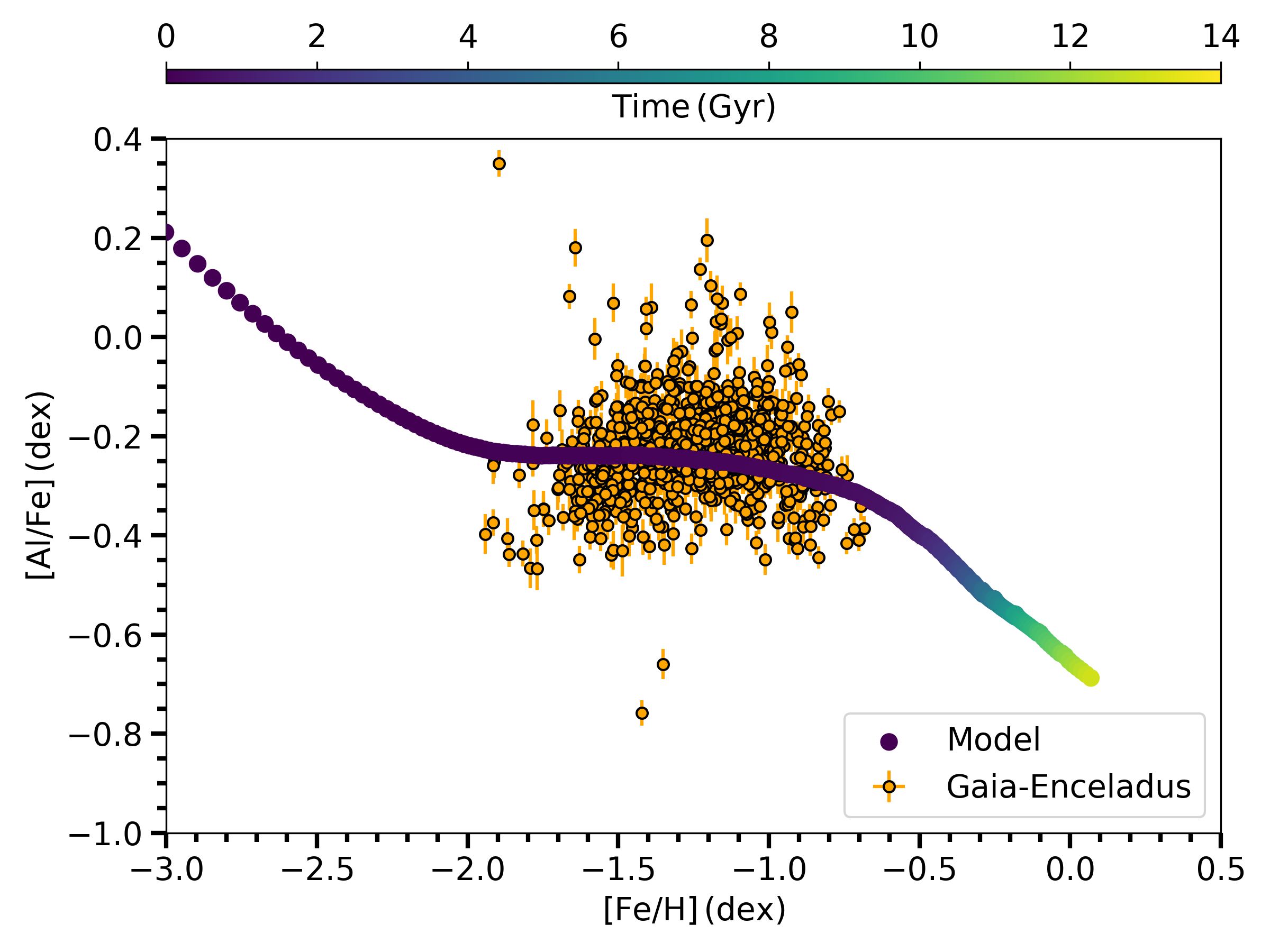}{0.30\textwidth}{}
               \fig{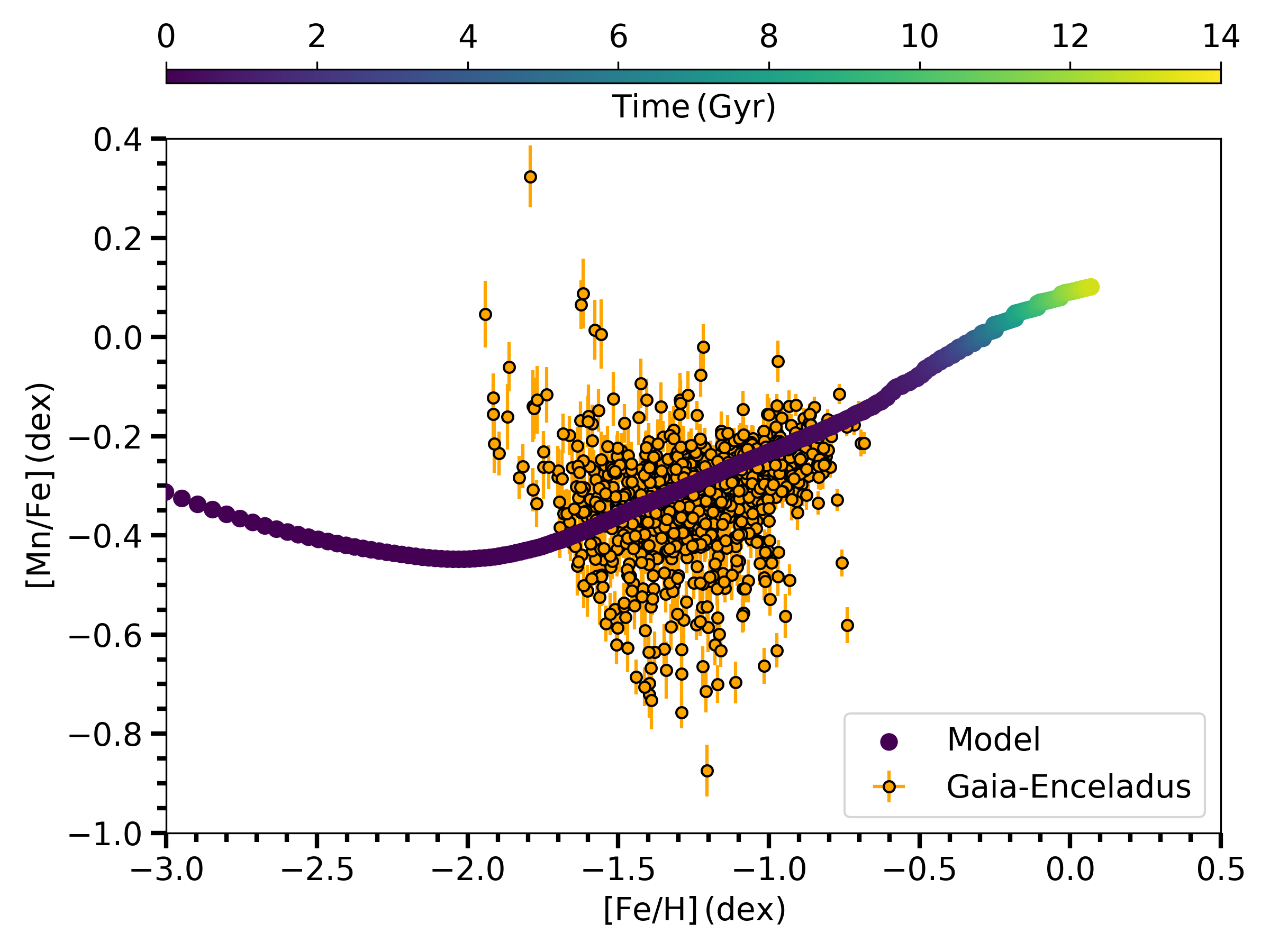}{0.30\textwidth}{}
               }\vspace{-1cm}
\caption{Predictions of the best model on different chemical planes.}\label{fig:four_element_model}
\end{figure*}

Although the model proposed to estimate the mass of the Gaia-Enceladus dwarf galaxy and understand its pre-merger chemical evolution shows satisfactory agreement in the $\mathrm{[Mg/Fe] \times [Fe/H]}$ plane, it is also essential to evaluate its predictions in other element planes to assess its overall reliability. Figure~\ref{fig:four_element_model} depicts the model’s predictions across various planes formed by Al, Mg, Mn, and Fe, along with the corresponding observational data. Even though the model was fitted exclusively in one particular plane, it exhibits equally good agreement with the observations in these other planes, capturing the general shape of the observational distributions. When observational uncertainties are considered, the model predictions prove remarkably consistent with the data. 

\begin{figure}
    \centering
    \includegraphics[width=\columnwidth]{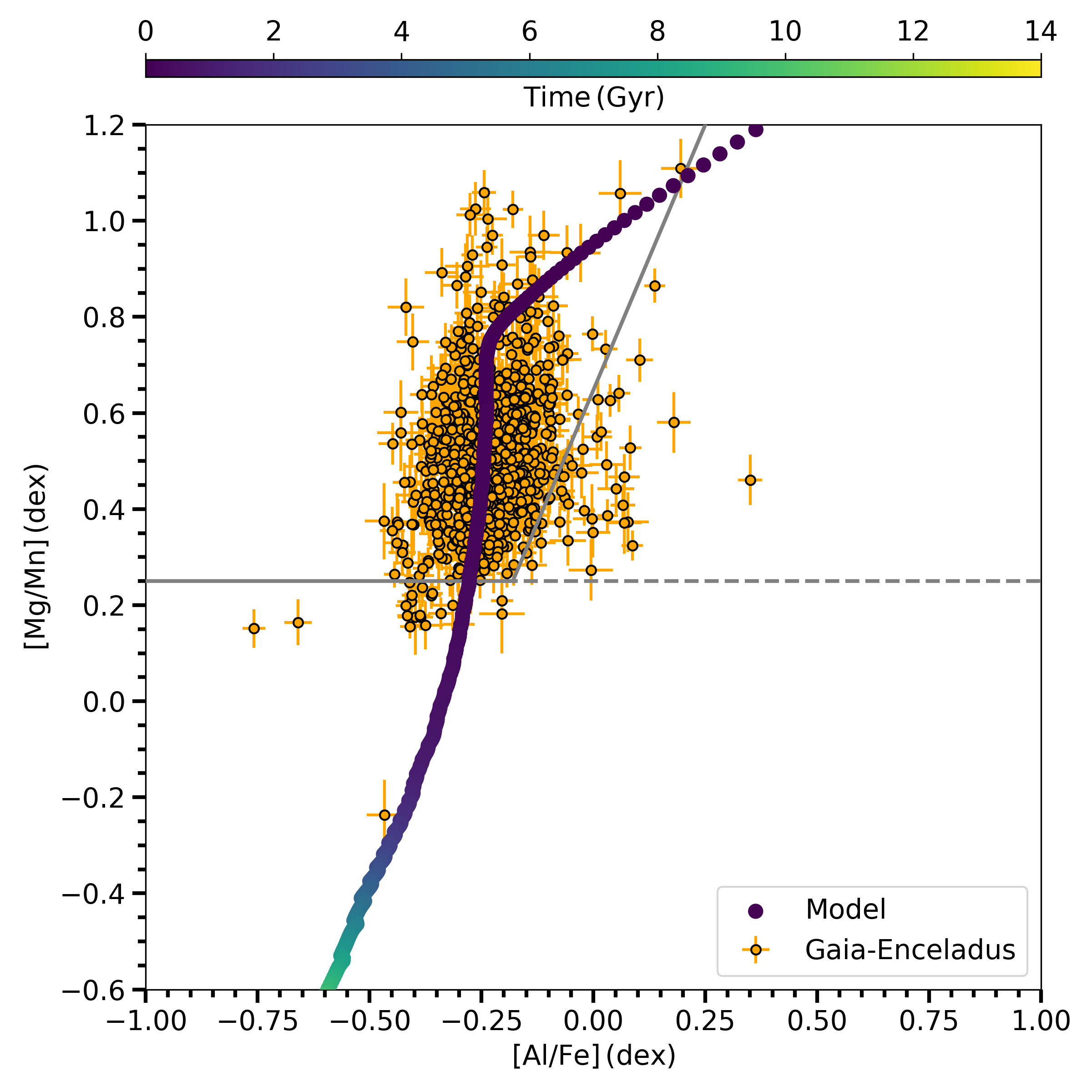}
    \caption{Chemical Evolution model predictions on the $\mathrm{[Mg/Mn]\times[Al/Fe]}$ plane.}
    \label{fig:alfe_mgmn_model}
\end{figure}

The same holds true for the $\mathrm{[Mg/Mn] \times [Al/Fe]}$ plane, which we employ in this study as one of our key diagnostic diagrams. Figure~\ref{fig:alfe_mgmn_model} shows that, in this plane, Gaia-Enceladus occupies the ex-situ region, indicating a chemical evolution pathway distinct from that of the Milky Way. Both the observational data and the model highlight this difference, which becomes even more apparent when compared with the solid grey lines proposed by \citet{Horta2023}. Overall, when examining all of these chemical planes, we find that—despite being fitted primarily to a single chemical plane—the model’s predictions in other planes remain in close agreement with the observational data.

\section{Discussion} \label{sec:discussion}

\subsection{Chemical Anomaly or Forgotten Merger? The Curious Case of Cluster 4}

Among the five clusters identified by our machine learning pipeline, Cluster 4, comprising 27 stars, does not exhibit chemo-dynamical properties consistent with any of the well-established accreted or in-situ Galactic populations reported in the literature (e.g., Gaia–Enceladus, Helmi, Sequoia, Sagittarius, Arjuna). Despite cross-validation using selection boundaries from recent studies \citep[e.g][]{Feuillet2020, Naidu2020, Horta2023}, this group remains unassociated with known merger remnants. We also evaluated the distribution of Cluster 4 in the action and velocity space diagrams presented in Figures \ref{fig:diagnostic_graphs}. The group occupies a compact region in the $J_R \times J_\phi$ plane, distinct from both the thick disk and Gaia–Enceladus, and appears dynamically hotter than typical in-situ stars. Notably, no strong clustering with other known accreted structures is evident.

Kinematically, stars in Cluster~4 are on extremely radial orbits, with a median eccentricity of $e_p = 0.88$, pericentric distance $R_{\mathrm{peri}} = 0.61$ kpc, and apocentric distance $R_{\mathrm{apo}} = 10.2$ kpc. The median angular momentum of the group is $L_z = -143\; \mathrm{kpc\,km\,s^{-1}}$, placing them on mildly retrograde orbits, while the median azimuthal velocity is $v_\phi= -2.2 \mathrm{km\,s^{-1}}$, indicative of a population that is neither strongly prograde nor retrograde. These orbital properties are incompatible with classical thick disk or canonical halo populations.

Chemically, the group is relatively metal-rich compared to classical Gaia–Enceladus stars, with a median $\mathrm{[Fe/H] = -0.70}$ dex. The stars also exhibit modest $\mathrm{[Mg/Fe] = 0.15}$ dex and low $\mathrm{[Mn/Fe] = -0.16}$, alongside near-solar $\mathrm{[Al/Fe]}$. The combined abundance ratio $\mathrm{[Mg/Mn] = 0.31}$ further positions this group outside typical accreted dwarf galaxy tracks. These patterns suggest either an unusual chemical enrichment history or a system that experienced early chemical self-enrichment but was later truncated.

We did not model the chemical evolution of Cluster~4 due to its small sample size and the lack of clear association with a known merger remnant. However, we note that its intermediate $\mathrm{[Mg/Fe]}$, enhanced $\mathrm{[Al/Fe]}$, and tight kinematics could be consistent with the debris of a minor merger or an early-formed in-situ population that has been dynamically heated. The absence of strong chemical coherence with major accreted populations strengthens the case for a distinct origin.

Taken together, the chemo-dynamical properties of Cluster~4 suggest that it may represent the relic of a previously unrecognized accretion event or a chemically distinct, dynamically heated in-situ population. While its intermediate $\mathrm{[Mg/Fe]}$, enhanced $\mathrm{[Al/Fe]}$, and compact orbital distribution support the possibility of a minor merger origin, the lack of strong chemical coherence with major accreted populations complicates a definitive classification. Among the known substructures in the literature, the properties of Cluster~4—particularly its distribution in energy, [Mg/Mn], [Fe/H], [Al/Fe], and [Mn/Fe]—show the strongest resemblance to the Eos structure recently characterized by \citet{Myeong2022}. While other known substructures display less compatibility across multiple abundance and dynamical planes, confirming a direct association with Eos remains speculative and would require further spectroscopic analysis and age-dating to substantiate this potential link.

\subsection{Comparision with Literature}\label{sec:comparision}

To contextualize our findings, we compared our results with previous studies that employed APOGEE data and chemical evolution models similar in scope to ours to derive constraints on Gaia–Enceladus. We begin this comparison with \citet{FernandezAlvar2018}, in which the authors analyzed halo stars by separating them into two distinct populations, akin to the approach of \citet{NissenSchuster1997}, based on their magnesium abundances—namely, high-Mg and low-Mg populations. The low-Mg population shows a strong agreement with our selected Gaia–Enceladus sample in terms of magnesium abundance (0.19 dex), although it corresponds to a more metal-poor population in terms of iron abundance. Given this, we infer that the majority of the stars selected in \citet{FernandezAlvar2018} likely belong to Gaia–Enceladus. Their proposed star formation rate (SFR) is consistent with our model, and the distribution in the $\mathrm{[Mg/Fe] \times [Fe/H]}$ plane also exhibits a similar trend. Overall, our results show good agreement with this study.

One of the most methodologically comparable studies in the recent literature is that of \citet{Vincenzo2019}. In this work, the authors model the chemical evolution of the Gaia-Enceladus dwarf galaxy using an exponential infall function similar to the one adopted in our study. However, the key difference between the two approaches lies in the treatment of the mass of the galaxy. While we treat the stellar mass of Gaia-Enceladus as a free parameter, \citet{Vincenzo2019} take the mass as a fixed value and leave other model parameters free instead. Moreover, in their analysis, the mass loading factor is not freed, but tested on three different values and fixed at 0.5. Despite these methodological differences, the results of \citet{Vincenzo2019} are in good agreement with ours. Specifically, they report a median iron abundance of $\mathrm{[Fe/H]} = -1.26$ dex for Gaia–Enceladus stars, which aligns well with the abundance distribution of our selected sample. In addition, the evolution of the star formation rate (SFR) in their model peaks around 0.5 Gyr, closely matching the temporal behavior inferred in our analysis, as illustrated in Figure~\ref{fig:sfr_sn_rate}. Taken together, these similarities underscore the consistency between the two studies in their predictions for the star formation history and chemical properties of Gaia–Enceladus.

In the study by \citet{Hasselquist2021}, the authors modeled the chemical evolution not only of Gaia–Enceladus but also of the LMC, SMC, Fornax, and Sagittarius dwarf galaxies using the FlexCE code \citep{Andrews2017}. These systems were analyzed in both the $\mathrm{[Mg/Fe] \times [Fe/H]}$ and $\mathrm{[Si/Fe] \times [Fe/H]}$ abundance planes. When comparing the results obtained in the $\mathrm{[Mg/Fe] \times [Fe/H]}$ plane to our findings, we observe that the chemical evolution model proposed by \citet{Hasselquist2021} shows an almost one-to-one agreement with our model predictions. Similarly, the evolution of the star formation rate (SFR) in their analysis peaks at a comparable epoch and exhibits a similar declining trend over time, indicating strong consistency between the two studies.

Examining the model parameters, we note that \citet{Hasselquist2021} fixed the collapse timescale at 2.5 Gyr, while our best-fit value was 3.30 Gyr—two values that are reasonably close. However, the mass-loading factor adopted in their model is nearly three times larger than ours, suggesting a more vigorous outflow and hence a more aggressive merger scenario than that inferred from our analysis. Despite these differences in modeling assumptions, both studies yield remarkably similar predictions in the $\mathrm{[Mg/Fe] \times [Fe/H]}$ plane, highlighting the robustness of the chemical evolution signatures and the overall compatibility of the results.

In \citet{Horta2021}, the authors modeled the chemical evolution of the Gaia–Enceladus dwarf galaxy in the $\mathrm{[Mg/Mn] \times [Al/Fe]}$ plane using the same model parameters adopted from \citet{Hasselquist2021}, implemented through the FlexCE framework. The results presented in Figure 2 of \citet{Horta2021} can be directly compared with our Figure~\ref{fig:alfe_mgmn_model}. A close inspection of these two panels reveals that the evolutionary tracks in both models begin at different initial points, pass through a turning point, and then diverge significantly thereafter. This divergence is primarily attributable to differences in the adopted yield sets. In particular, both models show notable discrepancies in their predictions for aluminium at low metallicities. These differences become especially pronounced in regions of the abundance plane where observational data are sparse or absent, while in the observationally constrained regions, the tracks tend to intersect. The sensitivity of chemical evolution models to the choice of stellar yields is well established in the literature and has been extensively discussed in \citet{Pepe2025}.

The fact that such discrepancies arise when applying the same model parameters from \citet{Hasselquist2021} to a different abundance plane, while the models exhibit near-perfect agreement in the $\mathrm{[Mg/Fe] \times [Fe/H]}$ plane, is particularly noteworthy. Moreover, the internal inconsistencies observed in the $\mathrm{[Si/Fe] \times [Fe/H]}$ plane in \citet{Hasselquist2021} further emphasize the point: chemical evolution models can yield markedly different outcomes when extrapolated across different abundance planes. This comparison demonstrates that chemically consistent trends can be reproduced under different, physically motivated parameter sets, underscoring the compatibility of our findings with the literature but also reflecting the inherent degeneracy of one-zone chemical evolution models.

\subsection{Chemical Evolution Scenario from Best Model Parameters} \label{sec:scenerio}

From our chemical evolution scenario, we infer that Gaia-Enceladus had a total baryonic mass of $4.93\times10^9\,M_\odot$, in good agreement with the broader literature range of $10^{8}$-$10^{10}\,M_\odot$ \citep{Helmi2018, Kruijssen2019, Vincenzo2019, Forbes2020, Limberg2022}. This best-fit mass, obtained via MCMC analysis of the ${\rm [Mg/Fe] \times [Fe/H]}$ plane, underpins a model in which Gaia-Enceladus experienced rapid gas depletion at the hands of the Milky Way, evidenced by a substantial mass-loading factor ($\eta\approx 2$). Such a high $\eta$ indicates vigorous outflows, thereby preventing the dwarf galaxy from retaining gas for extended periods.

Concurrently, the elevated star formation efficiency ($\epsilon$) with short star formation timescale ($\tau_\ast$) and short infall timescale ($\tau_\mathrm{infall}$)  imply that Gaia-Enceladus generated its stars swiftly, peaking roughly $0.5\,\mathrm{Gyr}$ after its formation. As shown in the Figure~\ref{fig:sfr_sn_rate}, these conditions are reflected in the supernova rates: while core-collapse (CC) supernovae display a single, rapid peak consistent with the burst of star formation, Type~Ia supernovae never dominate the enrichment process. This outcome not only explains the relatively low iron content (median $[\mathrm{Fe/H}]\approx -1.23$~dex) observed among Gaia-Enceladus stars but also supports the idea that prolonged SNe~Ia activity did not significantly elevate the iron abundance in this system. 

Overall, the scenario is consistent with a merger event in which Gaia-Enceladus rapidly formed and then lost a substantial portion of its gas through strong outflows, ultimately contributing a metal-poor stellar population and shaping the halo population we observe today in the Milky Way.

\subsubsection{Estimation Merger Time}

Determining the time of the Gaia–Enceladus merger is as crucial as estimating its initial mass for understanding the early chemical evolution and formation history of the Milky Way. However, the available observational constraints do not allow us to reliably infer the merger epoch directly from our model outputs. Nevertheless, under the assumption that star formation in the Gaia–Enceladus dwarf galaxy largely ceased following its accretion, we argue that our results provide sufficient indirect constraints to estimate the timing of the merger.

\begin{figure}
\includegraphics[width=\columnwidth]{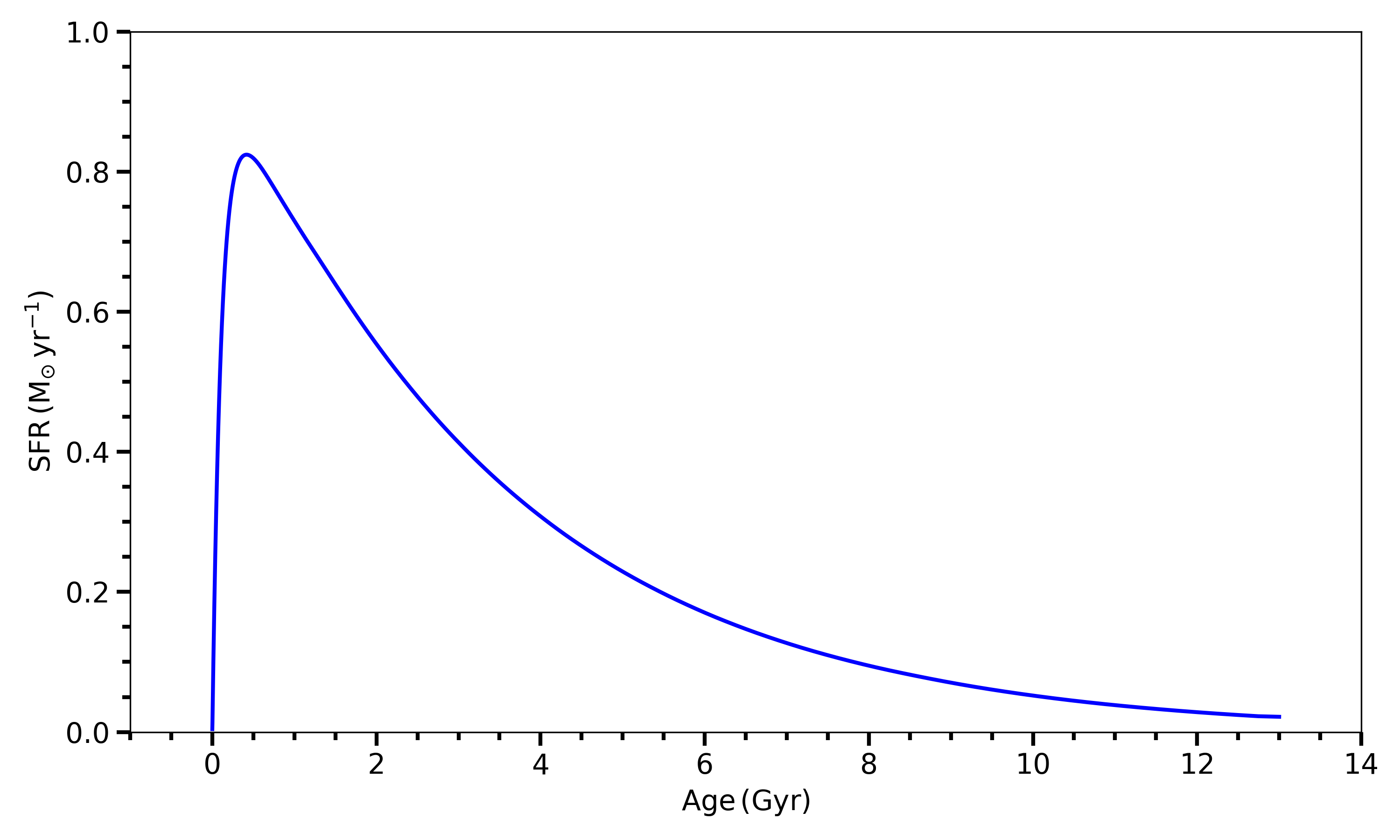}
\includegraphics[width=\columnwidth]{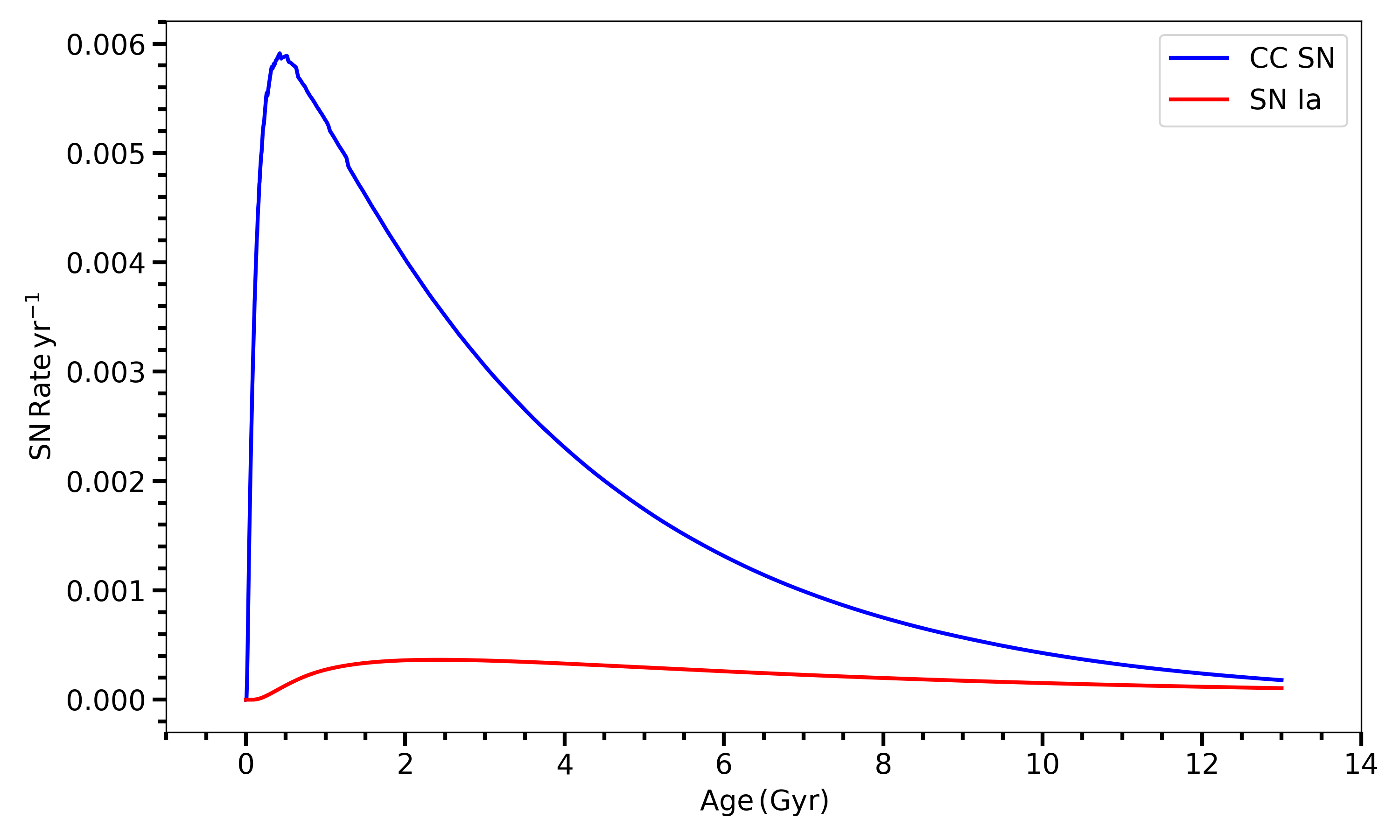}
\caption{Best model SFR and Supernova Rate over Time.}\label{fig:sfr_sn_rate}
\end{figure}

Under this assumption, inspection of Figure \ref{fig:sfr_sn_rate} reveals that star formation reaches a peak around 0.5 Gyr and subsequently enters a prolonged decline. The shape of this curve is governed by key model parameters—namely, $\tau_{\rm peak}$, $\tau_{\rm infall}$, and $\tau_\ast$—which are therefore essential for constraining the timing of the merger. The early peak of the star formation rate (SFR) near 0.5 Gyr reflects a rapid star formation phase driven by short infall and star formation timescales. The subsequent gradual decline, extending over nearly 4 Gyr, is primarily due to the longer infall timescale, estimated as $\tau_{\rm infall} = 3.30$ Gyr. Taking these timescales into account, and assuming that the Gaia–Enceladus merger effectively quenched star formation in the dwarf galaxy, we infer that the merger likely occurred within the first 4 Gyr of its evolution. This estimated timescale is consistent with values reported in previous studies \citep{Helmi2018, Vincenzo2019, Chaplin2020} that employed independent methods to determine the epoch of the merger.

Also, the peak value of the star formation rate (SFR) and the width of its distribution are consistent with the age distribution characteristics of Gaia-Enceladus stars reported in the literature. In particular, the peak around 0.5 Gyr corresponds well to the median age of 13.60$\pm$0.11 Gyr reported for Gaia-Enceladus stars by \citet{Limberg2022}, and is also in agreement with the broader age distribution range of 10–12 Gyr provided by \citet{Feuillet2021}. This agreement indicates that the SFR distribution predicted by the model aligns well with the age distribution properties determined for Gaia-Enceladus stars, thereby supporting the model’s capacity to accurately describe the chemical evolution of the galaxy.

\subsubsection{What if there was no Merger Event?}

\begin{table}[]
\caption{Chemical Evolution Model Best-Fit Results.}
\begin{tabular}{l|cc}
\hline
Paremeter                & Best-Fit         & Unit                  \\ \hline
$M_{\mathrm{gas}}$       & $4.93^{+0.32}_{-0.72}$ & $10^9 \times M_\odot$ \\
$\tau_{\mathrm{peak}}$   & $1.73^{+0.41}_{-0.39}$ & Gyr                   \\
$\tau_{\mathrm{infall}}$ & $3.30^{+0.38}_{-0.36}$ & Gyr                   \\
$\eta$                   & $1.93^{+0.04}_{-0.08}$ &                       \\
$\epsilon$               & $3.85^{+1.41}_{-0.71}$ &                       \\
$\tau_{\ast}$            & $1.18^{+0.42}_{-0.21}$ & Gyr   \\
\hline
\hline
\end{tabular}
\end{table}

One of the key questions addressed in this study is: “Where would the Gaia–Enceladus stars be located in chemical abundance planes if the merger had not occurred?” Although this question may appear speculative at first glance, it is in fact crucial for understanding how the chemical evolution of dwarf galaxies differs from that of spiral galaxies. Moreover, exploring the potential answers to this question provides insight into how the remmants of dwarf galaxies that merged after completing the majority of their chemical evolution can be distinguished from Milky Way stars in chemical abundance space.

To address this question, Figures \ref{fig:four_element_model} and \ref{fig:alfe_mgmn_model} should be considered together. In particular, Figure \ref{fig:alfe_mgmn_model} highlights how distinct the chemical evolution of the Gaia–Enceladus dwarf galaxy is from that of the Milky Way, especially when interpreted in light of the separation lines proposed by \citet{Horta2023}. The future evolution predicted by our model is located in the lower-left region of the diagram—below the horizontal line at $\mathrm{[Mg/Mn] = 0.25}$. Given that this plane effectively separates ex-situ from in-situ populations based on their chemical properties, it can be regarded as a principal diagnostic space in this study. Consequently, analyzing the merger-free trajectory in this plane becomes particularly significant. Notably, the model’s predicted future evolution falls within the region defined by \citet{Horta2023} as characteristic of in-situ stars. As shown in Figure \ref{fig:input_diagrams}, this region is only sparsely populated by stars in our sample.

This observation raises intriguing implications. If the Gaia–Enceladus dwarf galaxy had continued its chemical evolution undisturbed by Milky Way, it may have gradually evolved into regions of chemical space typically occupied by in-situ populations, despite its ex-situ origin. This convergence underscores the challenges inherent in disentangling stellar populations solely on the basis of chemical abundances, particularly at late times when abundance patterns can become degenerate. It also highlights the importance of merger timing: an earlier accretion event, such as that of Gaia–Enceladus, preserves the distinct chemical identity of the progenitor, whereas a later or slower merger might have resulted in a stellar population that is chemically indistinguishable from in-situ stars. Therefore, evaluating the counterfactual scenario—i.e., the absence of a merger—not only enhances our understanding of dwarf galaxy evolution but also provides a framework for interpreting the chemical footprints of accreted systems in the Galactic halo.

\subsection{Gaia-Enceladus Mass Estimation}

Our chemical evolution model introduces the initial gas mass of Gaia-Enceladus as a free parameter in the fitting procedure, distinguishing it from previous studies that estimated the stellar mass of Gaia-Enceladus based on various observational or theoretical assumptions. To enable a meaningful comparison, we adopt the standard assumption that the initial gas mass of the system is approximately equal to its final stellar mass—that is, $M_{\mathrm{gas}} \approx M_{\star}$. Previous estimates for the stellar mass of Gaia-Enceladus vary substantially: for example, \citet{Helmi2018} reported a mass of $6\times10^8\,M_{\odot}$, whereas \citet{Vincenzo2019} derived a significantly higher value of $\sim5\times10^9\,M_{\odot}$. \citet{Feuillet2020} estimated the progenitor mass to lie between $\sim10^{8.5}$ and $10^{9.5}\,M_{\odot}$, while \citet{Limberg2022} suggested a value of $\sim1.3\times10^9\,M_{\odot}$. In contrast, lower estimates were proposed by \citet{Kruijssen2019} and \citet{Forbes2020}, who inferred masses of $\sim3\times10^8\,M_{\odot}$ and $\sim8\times10^8\,M_{\odot}$, respectively, based on scaling relations and cosmological analogies. More recently, \citet{Lane2023} suggested a significantly lower mass of $1.45 \times 10^8\,M_\odot$.

In our analysis, by allowing the initial gas mass to emerge self-consistently from the best-fit chemical evolution solution, we find $M_{\mathrm{gas}}=4.93^{+0.32}_{-0.72}\times10^9\,M{\odot}$. This value not only aligns well with the higher end of existing estimates—such as that of \citet{Vincenzo2019}—but also remains broadly consistent with the wider range of values derived through diverse methods. Our chemical evolution model–based approach predicts a mass for Gaia-Enceladus that is comparable to or even exceeds that of the Large Magellanic Cloud, reinforcing the possibility that it was one of the most massive satellites ever accreted by the Milky Way, as previously suggested in the literature. These findings underscore the value of chemical evolution modeling as a robust tool for probing the merger history and initial mass scales of ancient dwarf galaxies.

\section{Summary and Conclusion}\label{sec:conclusion}

In summary, we analyzed multi-dimensional chemical and kinematic data using advanced machine learning techniques (e.g., t-SNE, HDBSCAN) to isolate Gaia–Enceladus members and characterize their distinct abundance patterns. Our results indicate a relatively massive initial system with a short, intense star formation history, followed by strong outflows and delayed Type Ia supernova contributions. This approach provides new insights into the early enrichment of the Milky Way halo and underscores the importance of robust yield prescriptions and numerical methods in Galactic archaeology.

Importantly, we show that clustering in chemo-dynamical space—when validated against entropy-based metrics and literature-defined structures—can efficiently recover Gaia–Enceladus. The model we present is not only consistent with observational trends in several chemical planes but also highlights the degeneracy between late-time chemical evolution and merger timing, suggesting that early accretion events leave more readily distinguishable chemical footprints.

Our MCMC-constrained chemical evolution model provides a statistically robust estimate for the initial gas mass of Gaia–Enceladus, yielding $M_{gas}=4.93_{-0.72}^{+0.32}\times10^9\,M_{\odot}$. This value is consistent with, and even exceeds, previous stellar mass estimates in the literature, placing Gaia–Enceladus among the most massive satellites ever accreted by the Milky Way. Furthermore, the inferred star formation history, characterized by rapid early activity and high mass-loading factors, supports the scenario of a chemically truncated system whose evolutionary path diverged sharply from that of the Milky Way.

Finally, we emphasize that combining unsupervised clustering with physically motivated chemical evolution models constitutes a powerful methodological framework. This synergy not only enhances the selection of merger debris in the halo but also enables rigorous quantitative reconstructions of their progenitor properties. Our chemically constrained estimate instead highlights Gaia-Enceladus as a gas-rich system, likely among the most massive dwarf galaxies ever accreted by the Milky Way. This result strengthens the case for chemical evolution modeling as a quantitative and independent pathway to reconstruct the pre-merger properties of ancient satellites and the assembly history of the Galactic halo.

\begin{acknowledgments}
We are especially grateful to the anonymous referee for their constructive feedback and thoughtful guidance, which significantly improved the clarity and robustness of the manuscript. We thank Danny Horta Darrington for his valuable insights and constructive discussions that contributed to the development of this work. We also extend our gratitude to all members of the Galactic Structure Research Group at Istanbul University for their continuous support and fruitful discussions throughout the course of this study. This study was partially supported by the Scientific and Technological Research Council (TÜBİTAK) MFAG-123F227. This study was funded by the Scientific Research Projects Coordination Unit of the Istanbul University. Project number: 40044, FBA-2023-39380 and FDK-2025-41537.
\end{acknowledgments}

%



\software{python \citep{Python}, \textit{OMEGA+}\citep{Cote2017,Cote2018}, astropy \citep{Astropy2013, Astropy2018},  numpy \citep{Numpy, Numpy2020}, scikit-learn \citep{scikit-learn}, matplotlib \citep{Matplotlib, Matplotlib2007}}




\bibliography{references}{}

\begin{thebibliography}{}
\expandafter\ifx\csname natexlab\endcsname\relax\def\natexlab#1{#1}\fi
\providecommand{\url}[1]{\href{#1}{#1}}
\providecommand{\dodoi}[1]{doi:~\href{http://doi.org/#1}{\nolinkurl{#1}}}
\providecommand{\doeprint}[1]{\href{http://ascl.net/#1}{\nolinkurl{http://ascl.net/#1}}}
\providecommand{\doarXiv}[1]{\href{https://arxiv.org/abs/#1}{\nolinkurl{https://arxiv.org/abs/#1}}}

\bibitem[{{Abdurro'uf} {et~al.}(2022){Abdurro'uf}, {Accetta}, {Aerts}, {Silva Aguirre}, {Ahumada}, {Ajgaonkar}, {Filiz Ak}, {Alam}, {Allende Prieto}, {Almeida}, {Anders}, {Anderson}, {Andrews}, {Anguiano}, {Aquino-Ort{\'\i}z}, {Arag{\'o}n-Salamanca}, {Argudo-Fern{\'a}ndez}, {Ata}, {Aubert}, {Avila-Reese}, {Badenes}, {Barb{\'a}}, {Barger}, {Barrera-Ballesteros}, {Beaton}, {Beers}, {Belfiore}, {Bender}, {Bernardi}, {Bershady}, {Beutler}, {Bidin}, {Bird}, {Bizyaev}, {Blanc}, {Blanton}, {Boardman}, {Bolton}, {Boquien}, {Borissova}, {Bovy}, {Brandt}, {Brown}, {Brownstein}, {Brusa}, {Buchner}, {Bundy}, {Burchett}, {Bureau}, {Burgasser}, {Cabang}, {Campbell}, {Cappellari}, {Carlberg}, {Wanderley}, {Carrera}, {Cash}, {Chen}, {Chen}, {Cherinka}, {Chiappini}, {Choi}, {Chojnowski}, {Chung}, {Clerc}, {Cohen}, {Comerford}, {Comparat}, {da Costa}, {Covey}, {Crane}, {Cruz-Gonzalez}, {Culhane}, {Cunha}, {Dai}, {Damke}, {Darling}, {Davidson}, {Davies}, {Dawson}, {De Lee}, {Diamond-Stanic}, {Cano-D{\'\i}az}, {S{\'a}nchez},
  {Donor}, {Duckworth}, {Dwelly}, {Eisenstein}, {Elsworth}, {Emsellem}, {Eracleous}, {Escoffier}, {Fan}, {Farr}, {Feng}, {Fern{\'a}ndez-Trincado}, {Feuillet}, {Filipp}, {Fillingham}, {Frinchaboy}, {Fromenteau}, {Galbany}, {Garc{\'\i}a}, {Garc{\'\i}a-Hern{\'a}ndez}, {Ge}, {Geisler}, {Gelfand}, {G{\'e}ron}, {Gibson}, {Goddy}, {Godoy-Rivera}, {Grabowski}, {Green}, {Greener}, {Grier}, {Griffith}, {Guo}, {Guy}, {Hadjara}, {Harding}, {Hasselquist}, {Hayes}, {Hearty}, {Hern{\'a}ndez}, {Hill}, {Hogg}, {Holtzman}, {Horta}, {Hsieh}, {Hsu}, {Hsu}, {Huber}, {Huertas-Company}, {Hutchinson}, {Hwang}, {Ibarra-Medel}, {Chitham}, {Ilha}, {Imig}, {Jaekle}, {Jayasinghe}, {Ji}, {Johnson}, {Jones}, {J{\"o}nsson}, {Katkov}, {Khalatyan}, {Kinemuchi}, {Kisku}, {Knapen}, {Kneib}, {Kollmeier}, {Kong}, {Kounkel}, {Kreckel}, {Krishnarao}, {Lacerna}, {Lane}, {Langgin}, {Lavender}, {Law}, {Lazarz}, {Leung}, {Leung}, {Lewis}, {Li}, {Li}, {Lian}, {Liang}, {Lin}, {Lin}, {Lin}, {Lintott}, {Long}, {Longa-Pe{\~n}a}, {L{\'o}pez-Cob{\'a}}, {Lu},
  {Lundgren}, {Luo}, {Mackereth}, {de la Macorra}, {Mahadevan}, {Majewski}, {Manchado}, {Mandeville}, {Maraston}, {Margalef-Bentabol}, {Masseron}, {Masters}, {Mathur}, {McDermid}, {Mckay}, {Merloni}, {Merrifield}, {Meszaros}, {Miglio}, {Di Mille}, {Minniti}, {Minsley}, {Monachesi}, {Moon}, {Mosser}, {Mulchaey}, {Muna}, {Mu{\~n}oz}, {Myers}, {Myers}, {Nadathur}, {Nair}, {Nandra}, {Neumann}, {Newman}, {Nidever}, {Nikakhtar}, {Nitschelm}, {O'Connell}, {Garma-Oehmichen}, {Luan Souza de Oliveira}, {Olney}, {Oravetz}, {Ortigoza-Urdaneta}, {Osorio}, {Otter}, {Pace}, {Padilla}, {Pan}, {Pan}, {Parikh}, {Parker}, {Peirani}, {Pe{\~n}a Ram{\'\i}rez}, {Penny}, {Percival}, {Perez-Fournon}, {Pinsonneault}, {Poidevin}, {Poovelil}, {Price-Whelan}, {B{\'a}rbara de Andrade Queiroz}, {Raddick}, {Ray}, {Rembold}, {Riddle}, {Riffel}, {Riffel}, {Rix}, {Robin}, {Rodr{\'\i}guez-Puebla}, {Roman-Lopes}, {Rom{\'a}n-Z{\'u}{\~n}iga}, {Rose}, {Ross}, {Rossi}, {Rubin}, {Salvato}, {S{\'a}nchez}, {S{\'a}nchez-Gallego}, {Sanderson}, {Santana
  Rojas}, {Sarceno}, {Sarmiento}, {Sayres}, {Sazonova}, {Schaefer}, {Schiavon}, {Schlegel}, {Schneider}, {Schultheis}, {Schwope}, {Serenelli}, {Serna}, {Shao}, {Shapiro}, {Sharma}, {Shen}, {Shetrone}, {Shu}, {Simon}, {Skrutskie}, {Smethurst}, {Smith}, {Sobeck}, {Spoo}, {Sprague}, {Stark}, {Stassun}, {Steinmetz}, {Stello}, {Stone-Martinez}, {Storchi-Bergmann}, {Stringfellow}, {Stutz}, {Su}, {Taghizadeh-Popp}, {Talbot}, {Tayar}, {Telles}, {Teske}, {Thakar}, {Theissen}, {Tkachenko}, {Thomas}, {Tojeiro}, {Hernandez Toledo}, {Troup}, {Trump}, {Trussler}, {Turner}, {Tuttle}, {Unda-Sanzana}, {V{\'a}zquez-Mata}, {Valentini}, {Valenzuela}, {Vargas-Gonz{\'a}lez}, {Vargas-Maga{\~n}a}, {Alfaro}, {Villanova}, {Vincenzo}, {Wake}, {Warfield}, {Washington}, {Weaver}, {Weijmans}, {Weinberg}, {Weiss}, {Westfall}, {Wild}, {Wilde}, {Wilson}, {Wilson}, {Wilson}, {Wolf}, {Wood-Vasey}, {Yan}, {Zamora}, {Zasowski}, {Zhang}, {Zhao}, {Zheng}, {Zheng}, \& {Zhu}}]{ApogeeDR17}
{Abdurro'uf}, {Accetta}, K., {Aerts}, C., {et~al.} 2022, \apjs, 259, 35, \dodoi{10.3847/1538-4365/ac4414}

\bibitem[{{Andrews} {et~al.}(2017){Andrews}, {Weinberg}, {Sch{\"o}nrich}, \& {Johnson}}]{Andrews2017}
{Andrews}, B.~H., {Weinberg}, D.~H., {Sch{\"o}nrich}, R., \& {Johnson}, J.~A. 2017, \apj, 835, 224, \dodoi{10.3847/1538-4357/835/2/224}

\bibitem[{{Asplund} {et~al.}(2009){Asplund}, {Grevesse}, {Sauval}, \& {Scott}}]{Asplund2009}
{Asplund}, M., {Grevesse}, N., {Sauval}, A.~J., \& {Scott}, P. 2009, \araa, 47, 481, \dodoi{10.1146/annurev.astro.46.060407.145222}

\bibitem[{{Astropy Collaboration} {et~al.}(2013){Astropy Collaboration}, {Robitaille}, {Tollerud}, {Greenfield}, {Droettboom}, {Bray}, {Aldcroft}, {Davis}, {Ginsburg}, {Price-Whelan}, {Kerzendorf}, {Conley}, {Crighton}, {Barbary}, {Muna}, {Ferguson}, {Grollier}, {Parikh}, {Nair}, {Unther}, {Deil}, {Woillez}, {Conseil}, {Kramer}, {Turner}, {Singer}, {Fox}, {Weaver}, {Zabalza}, {Edwards}, {Azalee Bostroem}, {Burke}, {Casey}, {Crawford}, {Dencheva}, {Ely}, {Jenness}, {Labrie}, {Lim}, {Pierfederici}, {Pontzen}, {Ptak}, {Refsdal}, {Servillat}, \& {Streicher}}]{Astropy2013}
{Astropy Collaboration}, {Robitaille}, T.~P., {Tollerud}, E.~J., {et~al.} 2013, \aap, 558, A33, \dodoi{10.1051/0004-6361/201322068}

\bibitem[{{Astropy Collaboration} {et~al.}(2018){Astropy Collaboration}, {Price-Whelan}, {Sip{\H{o}}cz}, {G{\"u}nther}, {Lim}, {Crawford}, {Conseil}, {Shupe}, {Craig}, {Dencheva}, {Ginsburg}, {VanderPlas}, {Bradley}, {P{\'e}rez-Su{\'a}rez}, {de Val-Borro}, {Aldcroft}, {Cruz}, {Robitaille}, {Tollerud}, {Ardelean}, {Babej}, {Bach}, {Bachetti}, {Bakanov}, {Bamford}, {Barentsen}, {Barmby}, {Baumbach}, {Berry}, {Biscani}, {Boquien}, {Bostroem}, {Bouma}, {Brammer}, {Bray}, {Breytenbach}, {Buddelmeijer}, {Burke}, {Calderone}, {Cano Rodr{\'\i}guez}, {Cara}, {Cardoso}, {Cheedella}, {Copin}, {Corrales}, {Crichton}, {D'Avella}, {Deil}, {Depagne}, {Dietrich}, {Donath}, {Droettboom}, {Earl}, {Erben}, {Fabbro}, {Ferreira}, {Finethy}, {Fox}, {Garrison}, {Gibbons}, {Goldstein}, {Gommers}, {Greco}, {Greenfield}, {Groener}, {Grollier}, {Hagen}, {Hirst}, {Homeier}, {Horton}, {Hosseinzadeh}, {Hu}, {Hunkeler}, {Ivezi{\'c}}, {Jain}, {Jenness}, {Kanarek}, {Kendrew}, {Kern}, {Kerzendorf}, {Khvalko}, {King}, {Kirkby}, {Kulkarni},
  {Kumar}, {Lee}, {Lenz}, {Littlefair}, {Ma}, {Macleod}, {Mastropietro}, {McCully}, {Montagnac}, {Morris}, {Mueller}, {Mumford}, {Muna}, {Murphy}, {Nelson}, {Nguyen}, {Ninan}, {N{\"o}the}, {Ogaz}, {Oh}, {Parejko}, {Parley}, {Pascual}, {Patil}, {Patil}, {Plunkett}, {Prochaska}, {Rastogi}, {Reddy Janga}, {Sabater}, {Sakurikar}, {Seifert}, {Sherbert}, {Sherwood-Taylor}, {Shih}, {Sick}, {Silbiger}, {Singanamalla}, {Singer}, {Sladen}, {Sooley}, {Sornarajah}, {Streicher}, {Teuben}, {Thomas}, {Tremblay}, {Turner}, {Terr{\'o}n}, {van Kerkwijk}, {de la Vega}, {Watkins}, {Weaver}, {Whitmore}, {Woillez}, {Zabalza}, \& {Astropy Contributors}}]{Astropy2018}
{Astropy Collaboration}, {Price-Whelan}, A.~M., {Sip{\H{o}}cz}, B.~M., {et~al.} 2018, \aj, 156, 123, \dodoi{10.3847/1538-3881/aabc4f}

\bibitem[{{Barb{\'a}} {et~al.}(2019){Barb{\'a}}, {Minniti}, {Geisler}, {Alonso-Garc{\'\i}a}, {Hempel}, {Monachesi}, {Arias}, \& {G{\'o}mez}}]{Barba2019}
{Barb{\'a}}, R.~H., {Minniti}, D., {Geisler}, D., {et~al.} 2019, \apjl, 870, L24, \dodoi{10.3847/2041-8213/aaf811}

\bibitem[{{Belokurov} {et~al.}(2018){Belokurov}, {Erkal}, {Evans}, {Koposov}, \& {Deason}}]{Belokurov2018}
{Belokurov}, V., {Erkal}, D., {Evans}, N.~W., {Koposov}, S.~E., \& {Deason}, A.~J. 2018, \mnras, 478, 611, \dodoi{10.1093/mnras/sty982}

\bibitem[{{Belokurov} \& {Kravtsov}(2024)}]{Belokurov2024}
{Belokurov}, V., \& {Kravtsov}, A. 2024, \mnras, 528, 3198, \dodoi{10.1093/mnras/stad3920}

\bibitem[{{Bland-Hawthorn} \& {Gerhard}(2016)}]{BlandHawthorn2016}
{Bland-Hawthorn}, J., \& {Gerhard}, O. 2016, \araa, 54, 529, \dodoi{10.1146/annurev-astro-081915-023441}

\bibitem[{{Bovy}(2015)}]{galpy}
{Bovy}, J. 2015, \apjs, 216, 29, \dodoi{10.1088/0067-0049/216/2/29}

\bibitem[{Carrillo {et~al.}(2023)Carrillo, Deason, Fattahi, Callingham, \& Grand}]{Carrillo2023}
Carrillo, A., Deason, A.~J., Fattahi, A., Callingham, T.~M., \& Grand, R. J.~J. 2023, Monthly Notices of the Royal Astronomical Society, 527, 2165, \dodoi{10.1093/mnras/stad3274}

\bibitem[{{Carrillo} {et~al.}(2024){Carrillo}, {Deason}, {Fattahi}, {Callingham}, \& {Grand}}]{Carrillo2024}
{Carrillo}, A., {Deason}, A.~J., {Fattahi}, A., {Callingham}, T.~M., \& {Grand}, R. J.~J. 2024, \mnras, 527, 2165, \dodoi{10.1093/mnras/stad3274}

\bibitem[{{Carrillo} {et~al.}(2022){Carrillo}, {Hawkins}, {Jofr{\'e}}, {de Brito Silva}, {Das}, \& {Lucey}}]{Carrillo2022}
{Carrillo}, A., {Hawkins}, K., {Jofr{\'e}}, P., {et~al.} 2022, \mnras, 513, 1557, \dodoi{10.1093/mnras/stac518}

\bibitem[{{Carrillo} {et~al.}(2023){Carrillo}, {Ness}, {Hawkins}, {Sanderson}, {Wang}, {Wetzel}, \& {Bellardini}}]{Carillo2023}
{Carrillo}, A., {Ness}, M.~K., {Hawkins}, K., {et~al.} 2023, \apj, 942, 35, \dodoi{10.3847/1538-4357/aca1c7}

\bibitem[{{Chaplin} {et~al.}(2020){Chaplin}, {Serenelli}, {Miglio}, {Morel}, {Mackereth}, {Vincenzo}, {Kjeldsen}, {Basu}, {Ball}, {Stokholm}, {Verma}, {Mosumgaard}, {Silva Aguirre}, {Mazumdar}, {Ranadive}, {Antia}, {Lebreton}, {Ong}, {Appourchaux}, {Bedding}, {Christensen-Dalsgaard}, {Creevey}, {Garc{\'\i}a}, {Handberg}, {Huber}, {Kawaler}, {Lund}, {Metcalfe}, {Stassun}, {Bazot}, {Beck}, {Bell}, {Bergemann}, {Buzasi}, {Benomar}, {Bossini}, {Bugnet}, {Campante}, {Orhan}, {Corsaro}, {Gonz{\'a}lez-Cuesta}, {Davies}, {Di Mauro}, {Egeland}, {Elsworth}, {Gaulme}, {Ghasemi}, {Guo}, {Hall}, {Hasanzadeh}, {Hekker}, {Howe}, {Jenkins}, {Jim{\'e}nez}, {Kiefer}, {Kuszlewicz}, {Kallinger}, {Latham}, {Lundkvist}, {Mathur}, {Montalb{\'a}n}, {Mosser}, {Bed{\'o}n}, {Nielsen}, {{\"O}rtel}, {Rendle}, {Ricker}, {Rodrigues}, {Roxburgh}, {Safari}, {Schofield}, {Seager}, {Smalley}, {Stello}, {Szab{\'o}}, {Tayar}, {Theme{\ss}l}, {Thomas}, {Vanderspek}, {van Rossem}, {Vrard}, {Weiss}, {White}, {Winn}, \& {Y{\i}ld{\i}z}}]{Chaplin2020}
{Chaplin}, W.~J., {Serenelli}, A.~M., {Miglio}, A., {et~al.} 2020, Nature Astronomy, 4, 382, \dodoi{10.1038/s41550-019-0975-9}

\bibitem[{{Chiappini} {et~al.}(1997){Chiappini}, {Matteucci}, \& {Gratton}}]{Chiappini1997}
{Chiappini}, C., {Matteucci}, F., \& {Gratton}, R. 1997, \apj, 477, 765, \dodoi{10.1086/303726}

\bibitem[{{Conroy} {et~al.}(2019){Conroy}, {Bonaca}, {Cargile}, {Johnson}, {Caldwell}, {Naidu}, {Zaritsky}, {Fabricant}, {Moran}, {Rhee}, {Szentgyorgyi}, {Berlind}, {Calkins}, {Kattner}, \& {Ly}}]{H3survey}
{Conroy}, C., {Bonaca}, A., {Cargile}, P., {et~al.} 2019, \apj, 883, 107, \dodoi{10.3847/1538-4357/ab38b8}

\bibitem[{{C{\^o}t{\'e}} {et~al.}(2017){C{\^o}t{\'e}}, {O'Shea}, {Ritter}, {Herwig}, \& {Venn}}]{Cote2017}
{C{\^o}t{\'e}}, B., {O'Shea}, B.~W., {Ritter}, C., {Herwig}, F., \& {Venn}, K.~A. 2017, \apj, 835, 128, \dodoi{10.3847/1538-4357/835/2/128}

\bibitem[{{C{\^o}t{\'e}} {et~al.}(2018){C{\^o}t{\'e}}, {Silvia}, {O'Shea}, {Smith}, \& {Wise}}]{Cote2018}
{C{\^o}t{\'e}}, B., {Silvia}, D.~W., {O'Shea}, B.~W., {Smith}, B., \& {Wise}, J.~H. 2018, \apj, 859, 67, \dodoi{10.3847/1538-4357/aabe8f}

\bibitem[{{Das} {et~al.}(2020){Das}, {Hawkins}, \& {Jofr{\'e}}}]{Das2020}
{Das}, P., {Hawkins}, K., \& {Jofr{\'e}}, P. 2020, \mnras, 493, 5195, \dodoi{10.1093/mnras/stz3537}

\bibitem[{{De Silva} {et~al.}(2015){De Silva}, {Freeman}, {Bland-Hawthorn}, {Martell}, {de Boer}, {Asplund}, {Keller}, {Sharma}, {Zucker}, {Zwitter}, {Anguiano}, {Bacigalupo}, {Bayliss}, {Beavis}, {Bergemann}, {Campbell}, {Cannon}, {Carollo}, {Casagrande}, {Casey}, {Da Costa}, {D'Orazi}, {Dotter}, {Duong}, {Heger}, {Ireland}, {Kafle}, {Kos}, {Lattanzio}, {Lewis}, {Lin}, {Lind}, {Munari}, {Nataf}, {O'Toole}, {Parker}, {Reid}, {Schlesinger}, {Sheinis}, {Simpson}, {Stello}, {Ting}, {Traven}, {Watson}, {Wittenmyer}, {Yong}, \& {{\v{Z}}erjal}}]{Galah}
{De Silva}, G.~M., {Freeman}, K.~C., {Bland-Hawthorn}, J., {et~al.} 2015, \mnras, 449, 2604, \dodoi{10.1093/mnras/stv327}

\bibitem[{{Donlon} \& {Newberg}(2023)}]{Donlon2023}
{Donlon}, T., \& {Newberg}, H.~J. 2023, \apj, 944, 169, \dodoi{10.3847/1538-4357/acb150}

\bibitem[{{Fernandes} {et~al.}(2023){Fernandes}, {Mason}, {Horta}, {Schiavon}, {Hayes}, {Hasselquist}, {Feuillet}, {Beaton}, {J{\"o}nsson}, {Kisku}, {Lacerna}, {Lian}, {Minniti}, \& {Villanova}}]{Fernandes2023}
{Fernandes}, L., {Mason}, A.~C., {Horta}, D., {et~al.} 2023, \mnras, 519, 3611, \dodoi{10.1093/mnras/stac3543}

\bibitem[{{Fern{\'a}ndez-Alvar} {et~al.}(2018){Fern{\'a}ndez-Alvar}, {Carigi}, {Schuster}, {Hayes}, {{\'A}vila-Vergara}, {Majewski}, {Allende Prieto}, {Beers}, {S{\'a}nchez}, {Zamora}, {Garc{\'\i}a-Hern{\'a}ndez}, {Tang}, {Fern{\'a}ndez-Trincado}, {Tissera}, {Geisler}, \& {Villanova}}]{FernandezAlvar2018}
{Fern{\'a}ndez-Alvar}, E., {Carigi}, L., {Schuster}, W.~J., {et~al.} 2018, \apj, 852, 50, \dodoi{10.3847/1538-4357/aa9ced}

\bibitem[{Feuillet {et~al.}(2020)Feuillet, Feltzing, Sahlholdt, \& Casagrande}]{Feuillet2020}
Feuillet, D.~K., Feltzing, S., Sahlholdt, C.~L., \& Casagrande, L. 2020, Monthly Notices of the Royal Astronomical Society, 497, 109, \dodoi{10.1093/mnras/staa1888}

\bibitem[{{Feuillet} {et~al.}(2021){Feuillet}, {Sahlholdt}, {Feltzing}, \& {Casagrande}}]{Feuillet2021}
{Feuillet}, D.~K., {Sahlholdt}, C.~L., {Feltzing}, S., \& {Casagrande}, L. 2021, \mnras, 508, 1489, \dodoi{10.1093/mnras/stab2614}

\bibitem[{{Forbes}(2020)}]{Forbes2020}
{Forbes}, D.~A. 2020, \mnras, 493, 847, \dodoi{10.1093/mnras/staa245}

\bibitem[{{Freeman} \& {Bland-Hawthorn}(2002)}]{Freeman2002}
{Freeman}, K., \& {Bland-Hawthorn}, J. 2002, \araa, 40, 487, \dodoi{10.1146/annurev.astro.40.060401.093840}

\bibitem[{{Gaia Collaboration} {et~al.}(2016){Gaia Collaboration}, {Prusti}, {de Bruijne}, {Brown}, {Vallenari}, {Babusiaux}, {Bailer-Jones}, {Bastian}, {Biermann}, {Evans}, {Eyer}, {Jansen}, {Jordi}, {Klioner}, {Lammers}, {Lindegren}, {Luri}, {Mignard}, {Milligan}, {Panem}, {Poinsignon}, {Pourbaix}, {Randich}, {Sarri}, {Sartoretti}, {Siddiqui}, {Soubiran}, {Valette}, {van Leeuwen}, {Walton}, {Aerts}, {Arenou}, {Cropper}, {Drimmel}, {H{\o}g}, {Katz}, {Lattanzi}, {O'Mullane}, {Grebel}, {Holland}, {Huc}, {Passot}, {Bramante}, {Cacciari}, {Casta{\~n}eda}, {Chaoul}, {Cheek}, {De Angeli}, {Fabricius}, {Guerra}, {Hern{\'a}ndez}, {Jean-Antoine-Piccolo}, {Masana}, {Messineo}, {Mowlavi}, {Nienartowicz}, {Ord{\'o}{\~n}ez-Blanco}, {Panuzzo}, {Portell}, {Richards}, {Riello}, {Seabroke}, {Tanga}, {Th{\'e}venin}, {Torra}, {Els}, {Gracia-Abril}, {Comoretto}, {Garcia-Reinaldos}, {Lock}, {Mercier}, {Altmann}, {Andrae}, {Astraatmadja}, {Bellas-Velidis}, {Benson}, {Berthier}, {Blomme}, {Busso}, {Carry}, {Cellino}, {Clementini},
  {Cowell}, {Creevey}, {Cuypers}, {Davidson}, {De Ridder}, {de Torres}, {Delchambre}, {Dell'Oro}, {Ducourant}, {Fr{\'e}mat}, {Garc{\'\i}a-Torres}, {Gosset}, {Halbwachs}, {Hambly}, {Harrison}, {Hauser}, {Hestroffer}, {Hodgkin}, {Huckle}, {Hutton}, {Jasniewicz}, {Jordan}, {Kontizas}, {Korn}, {Lanzafame}, {Manteiga}, {Moitinho}, {Muinonen}, {Osinde}, {Pancino}, {Pauwels}, {Petit}, {Recio-Blanco}, {Robin}, {Sarro}, {Siopis}, {Smith}, {Smith}, {Sozzetti}, {Thuillot}, {van Reeven}, {Viala}, {Abbas}, {Abreu Aramburu}, {Accart}, {Aguado}, {Allan}, {Allasia}, {Altavilla}, {{\'A}lvarez}, {Alves}, {Anderson}, {Andrei}, {Anglada Varela}, {Antiche}, {Antoja}, {Ant{\'o}n}, {Arcay}, {Atzei}, {Ayache}, {Bach}, {Baker}, {Balaguer-N{\'u}{\~n}ez}, {Barache}, {Barata}, {Barbier}, {Barblan}, {Baroni}, {Barrado y Navascu{\'e}s}, {Barros}, {Barstow}, {Becciani}, {Bellazzini}, {Bellei}, {Bello Garc{\'\i}a}, {Belokurov}, {Bendjoya}, {Berihuete}, {Bianchi}, {Bienaym{\'e}}, {Billebaud}, {Blagorodnova}, {Blanco-Cuaresma}, {Boch},
  {Bombrun}, {Borrachero}, {Bouquillon}, {Bourda}, {Bouy}, {Bragaglia}, {Breddels}, {Brouillet}, {Br{\"u}semeister}, {Bucciarelli}, {Budnik}, {Burgess}, {Burgon}, {Burlacu}, {Busonero}, {Buzzi}, {Caffau}, {Cambras}, {Campbell}, {Cancelliere}, {Cantat-Gaudin}, {Carlucci}, {Carrasco}, {Castellani}, {Charlot}, {Charnas}, {Charvet}, {Chassat}, {Chiavassa}, {Clotet}, {Cocozza}, {Collins}, {Collins}, {Costigan}, {Crifo}, {Cross}, {Crosta}, {Crowley}, {Dafonte}, {Damerdji}, {Dapergolas}, {David}, {David}, {De Cat}, {de Felice}, {de Laverny}, {De Luise}, {De March}, {de Martino}, {de Souza}, {Debosscher}, {del Pozo}, {Delbo}, {Delgado}, {Delgado}, {di Marco}, {Di Matteo}, {Diakite}, {Distefano}, {Dolding}, {Dos Anjos}, {Drazinos}, {Dur{\'a}n}, {Dzigan}, {Ecale}, {Edvardsson}, {Enke}, {Erdmann}, {Escolar}, {Espina}, {Evans}, {Eynard Bontemps}, {Fabre}, {Fabrizio}, {Faigler}, {Falc{\~a}o}, {Farr{\`a}s Casas}, {Faye}, {Federici}, {Fedorets}, {Fern{\'a}ndez-Hern{\'a}ndez}, {Fernique}, {Fienga}, {Figueras}, {Filippi},
  {Findeisen}, {Fonti}, {Fouesneau}, {Fraile}, {Fraser}, {Fuchs}, {Furnell}, {Gai}, {Galleti}, {Galluccio}, {Garabato}, {Garc{\'\i}a-Sedano}, {Gar{\'e}}, {Garofalo}, {Garralda}, {Gavras}, {Gerssen}, {Geyer}, {Gilmore}, {Girona}, {Giuffrida}, {Gomes}, {Gonz{\'a}lez-Marcos}, {Gonz{\'a}lez-N{\'u}{\~n}ez}, {Gonz{\'a}lez-Vidal}, {Granvik}, {Guerrier}, {Guillout}, {Guiraud}, {G{\'u}rpide}, {Guti{\'e}rrez-S{\'a}nchez}, {Guy}, {Haigron}, {Hatzidimitriou}, {Haywood}, {Heiter}, {Helmi}, {Hobbs}, {Hofmann}, {Holl}, {Holland}, {Hunt}, {Hypki}, {Icardi}, {Irwin}, {Jevardat de Fombelle}, {Jofr{\'e}}, {Jonker}, {Jorissen}, {Julbe}, {Karampelas}, {Kochoska}, {Kohley}, {Kolenberg}, {Kontizas}, {Koposov}, {Kordopatis}, {Koubsky}, {Kowalczyk}, {Krone-Martins}, {Kudryashova}, {Kull}, {Bachchan}, {Lacoste-Seris}, {Lanza}, {Lavigne}, {Le Poncin-Lafitte}, {Lebreton}, {Lebzelter}, {Leccia}, {Leclerc}, {Lecoeur-Taibi}, {Lemaitre}, {Lenhardt}, {Leroux}, {Liao}, {Licata}, {Lindstr{\o}m}, {Lister}, {Livanou}, {Lobel}, {L{\"o}ffler},
  {L{\'o}pez}, {Lopez-Lozano}, {Lorenz}, {Loureiro}, {MacDonald}, {Magalh{\~a}es Fernandes}, {Managau}, {Mann}, {Mantelet}, {Marchal}, {Marchant}, {Marconi}, {Marie}, {Marinoni}, {Marrese}, {Marschalk{\'o}}, {Marshall}, {Mart{\'\i}n-Fleitas}, {Martino}, {Mary}, {Matijevi{\v{c}}}, {Mazeh}, {McMillan}, {Messina}, {Mestre}, {Michalik}, {Millar}, {Miranda}, {Molina}, {Molinaro}, {Molinaro}, {Moln{\'a}r}, {Moniez}, {Montegriffo}, {Monteiro}, {Mor}, {Mora}, {Morbidelli}, {Morel}, {Morgenthaler}, {Morley}, {Morris}, {Mulone}, {Muraveva}, {Musella}, {Narbonne}, {Nelemans}, {Nicastro}, {Noval}, {Ord{\'e}novic}, {Ordieres-Mer{\'e}}, {Osborne}, {Pagani}, {Pagano}, {Pailler}, {Palacin}, {Palaversa}, {Parsons}, {Paulsen}, {Pecoraro}, {Pedrosa}, {Pentik{\"a}inen}, {Pereira}, {Pichon}, {Piersimoni}, {Pineau}, {Plachy}, {Plum}, {Poujoulet}, {Pr{\v{s}}a}, {Pulone}, {Ragaini}, {Rago}, {Rambaux}, {Ramos-Lerate}, {Ranalli}, {Rauw}, {Read}, {Regibo}, {Renk}, {Reyl{\'e}}, {Ribeiro}, {Rimoldini}, {Ripepi}, {Riva}, {Rixon},
  {Roelens}, {Romero-G{\'o}mez}, {Rowell}, {Royer}, {Rudolph}, {Ruiz-Dern}, {Sadowski}, {Sagrist{\`a} Sell{\'e}s}, {Sahlmann}, {Salgado}, {Salguero}, {Sarasso}, {Savietto}, {Schnorhk}, {Schultheis}, {Sciacca}, {Segol}, {Segovia}, {Segransan}, {Serpell}, {Shih}, {Smareglia}, {Smart}, {Smith}, {Solano}, {Solitro}, {Sordo}, {Soria Nieto}, {Souchay}, {Spagna}, {Spoto}, {Stampa}, {Steele}, {Steidelm{\"u}ller}, {Stephenson}, {Stoev}, {Suess}, {S{\"u}veges}, {Surdej}, {Szabados}, {Szegedi-Elek}, {Tapiador}, {Taris}, {Tauran}, {Taylor}, {Teixeira}, {Terrett}, {Tingley}, {Trager}, {Turon}, {Ulla}, {Utrilla}, {Valentini}, {van Elteren}, {Van Hemelryck}, {van Leeuwen}, {Varadi}, {Vecchiato}, {Veljanoski}, {Via}, {Vicente}, {Vogt}, {Voss}, {Votruba}, {Voutsinas}, {Walmsley}, {Weiler}, {Weingrill}, {Werner}, {Wevers}, {Whitehead}, {Wyrzykowski}, {Yoldas}, {{\v{Z}}erjal}, {Zucker}, {Zurbach}, {Zwitter}, {Alecu}, {Allen}, {Allende Prieto}, {Amorim}, {Anglada-Escud{\'e}}, {Arsenijevic}, {Azaz}, {Balm}, {Beck}, {Bernstein},
  {Bigot}, {Bijaoui}, {Blasco}, {Bonfigli}, {Bono}, {Boudreault}, {Bressan}, {Brown}, {Brunet}, {Bunclark}, {Buonanno}, {Butkevich}, {Carret}, {Carrion}, {Chemin}, {Ch{\'e}reau}, {Corcione}, {Darmigny}, {de Boer}, {de Teodoro}, {de Zeeuw}, {Delle Luche}, {Domingues}, {Dubath}, {Fodor}, {Fr{\'e}zouls}, {Fries}, {Fustes}, {Fyfe}, {Gallardo}, {Gallegos}, {Gardiol}, {Gebran}, {Gomboc}, {G{\'o}mez}, {Grux}, {Gueguen}, {Heyrovsky}, {Hoar}, {Iannicola}, {Isasi Parache}, {Janotto}, {Joliet}, {Jonckheere}, {Keil}, {Kim}, {Klagyivik}, {Klar}, {Knude}, {Kochukhov}, {Kolka}, {Kos}, {Kutka}, {Lainey}, {LeBouquin}, {Liu}, {Loreggia}, {Makarov}, {Marseille}, {Martayan}, {Martinez-Rubi}, {Massart}, {Meynadier}, {Mignot}, {Munari}, {Nguyen}, {Nordlander}, {Ocvirk}, {O'Flaherty}, {Olias Sanz}, {Ortiz}, {Osorio}, {Oszkiewicz}, {Ouzounis}, {Palmer}, {Park}, {Pasquato}, {Peltzer}, {Peralta}, {P{\'e}turaud}, {Pieniluoma}, {Pigozzi}, {Poels}, {Prat}, {Prod'homme}, {Raison}, {Rebordao}, {Risquez}, {Rocca-Volmerange}, {Rosen},
  {Ruiz-Fuertes}, {Russo}, {Sembay}, {Serraller Vizcaino}, {Short}, {Siebert}, {Silva}, {Sinachopoulos}, {Slezak}, {Soffel}, {Sosnowska}, {Strai{\v{z}}ys}, {ter Linden}, {Terrell}, {Theil}, {Tiede}, {Troisi}, {Tsalmantza}, {Tur}, {Vaccari}, {Vachier}, {Valles}, {Van Hamme}, {Veltz}, {Virtanen}, {Wallut}, {Wichmann}, {Wilkinson}, {Ziaeepour}, \& {Zschocke}}]{Gaia16}
{Gaia Collaboration}, {Prusti}, T., {de Bruijne}, J.~H.~J., {et~al.} 2016, \aap, 595, A1, \dodoi{10.1051/0004-6361/201629272}

\bibitem[{{Gaia Collaboration} {et~al.}(2023){Gaia Collaboration}, {Vallenari}, {Brown}, {Prusti}, {de Bruijne}, {Arenou}, {Babusiaux}, {Biermann}, {Creevey}, {Ducourant}, {Evans}, {Eyer}, {Guerra}, {Hutton}, {Jordi}, {Klioner}, {Lammers}, {Lindegren}, {Luri}, {Mignard}, {Panem}, {Pourbaix}, {Randich}, {Sartoretti}, {Soubiran}, {Tanga}, {Walton}, {Bailer-Jones}, {Bastian}, {Drimmel}, {Jansen}, {Katz}, {Lattanzi}, {van Leeuwen}, {Bakker}, {Cacciari}, {Casta{\~n}eda}, {De Angeli}, {Fabricius}, {Fouesneau}, {Fr{\'e}mat}, {Galluccio}, {Guerrier}, {Heiter}, {Masana}, {Messineo}, {Mowlavi}, {Nicolas}, {Nienartowicz}, {Pailler}, {Panuzzo}, {Riclet}, {Roux}, {Seabroke}, {Sordo}, {Th{\'e}venin}, {Gracia-Abril}, {Portell}, {Teyssier}, {Altmann}, {Andrae}, {Audard}, {Bellas-Velidis}, {Benson}, {Berthier}, {Blomme}, {Burgess}, {Busonero}, {Busso}, {C{\'a}novas}, {Carry}, {Cellino}, {Cheek}, {Clementini}, {Damerdji}, {Davidson}, {de Teodoro}, {Nu{\~n}ez Campos}, {Delchambre}, {Dell'Oro}, {Esquej},
  {Fern{\'a}ndez-Hern{\'a}ndez}, {Fraile}, {Garabato}, {Garc{\'\i}a-Lario}, {Gosset}, {Haigron}, {Halbwachs}, {Hambly}, {Harrison}, {Hern{\'a}ndez}, {Hestroffer}, {Hodgkin}, {Holl}, {Jan{\ss}en}, {Jevardat de Fombelle}, {Jordan}, {Krone-Martins}, {Lanzafame}, {L{\"o}ffler}, {Marchal}, {Marrese}, {Moitinho}, {Muinonen}, {Osborne}, {Pancino}, {Pauwels}, {Recio-Blanco}, {Reyl{\'e}}, {Riello}, {Rimoldini}, {Roegiers}, {Rybizki}, {Sarro}, {Siopis}, {Smith}, {Sozzetti}, {Utrilla}, {van Leeuwen}, {Abbas}, {{\'A}brah{\'a}m}, {Abreu Aramburu}, {Aerts}, {Aguado}, {Ajaj}, {Aldea-Montero}, {Altavilla}, {{\'A}lvarez}, {Alves}, {Anders}, {Anderson}, {Anglada Varela}, {Antoja}, {Baines}, {Baker}, {Balaguer-N{\'u}{\~n}ez}, {Balbinot}, {Balog}, {Barache}, {Barbato}, {Barros}, {Barstow}, {Bartolom{\'e}}, {Bassilana}, {Bauchet}, {Becciani}, {Bellazzini}, {Berihuete}, {Bernet}, {Bertone}, {Bianchi}, {Binnenfeld}, {Blanco-Cuaresma}, {Blazere}, {Boch}, {Bombrun}, {Bossini}, {Bouquillon}, {Bragaglia}, {Bramante}, {Breedt},
  {Bressan}, {Brouillet}, {Brugaletta}, {Bucciarelli}, {Burlacu}, {Butkevich}, {Buzzi}, {Caffau}, {Cancelliere}, {Cantat-Gaudin}, {Carballo}, {Carlucci}, {Carnerero}, {Carrasco}, {Casamiquela}, {Castellani}, {Castro-Ginard}, {Chaoul}, {Charlot}, {Chemin}, {Chiaramida}, {Chiavassa}, {Chornay}, {Comoretto}, {Contursi}, {Cooper}, {Cornez}, {Cowell}, {Crifo}, {Cropper}, {Crosta}, {Crowley}, {Dafonte}, {Dapergolas}, {David}, {David}, {de Laverny}, {De Luise}, {De March}, {De Ridder}, {de Souza}, {de Torres}, {del Peloso}, {del Pozo}, {Delbo}, {Delgado}, {Delisle}, {Demouchy}, {Dharmawardena}, {Di Matteo}, {Diakite}, {Diener}, {Distefano}, {Dolding}, {Edvardsson}, {Enke}, {Fabre}, {Fabrizio}, {Faigler}, {Fedorets}, {Fernique}, {Fienga}, {Figueras}, {Fournier}, {Fouron}, {Fragkoudi}, {Gai}, {Garcia-Gutierrez}, {Garcia-Reinaldos}, {Garc{\'\i}a-Torres}, {Garofalo}, {Gavel}, {Gavras}, {Gerlach}, {Geyer}, {Giacobbe}, {Gilmore}, {Girona}, {Giuffrida}, {Gomel}, {Gomez}, {Gonz{\'a}lez-N{\'u}{\~n}ez},
  {Gonz{\'a}lez-Santamar{\'\i}a}, {Gonz{\'a}lez-Vidal}, {Granvik}, {Guillout}, {Guiraud}, {Guti{\'e}rrez-S{\'a}nchez}, {Guy}, {Hatzidimitriou}, {Hauser}, {Haywood}, {Helmer}, {Helmi}, {Sarmiento}, {Hidalgo}, {Hilger}, {H{\l}adczuk}, {Hobbs}, {Holland}, {Huckle}, {Jardine}, {Jasniewicz}, {Jean-Antoine Piccolo}, {Jim{\'e}nez-Arranz}, {Jorissen}, {Juaristi Campillo}, {Julbe}, {Karbevska}, {Kervella}, {Khanna}, {Kontizas}, {Kordopatis}, {Korn}, {K{\'o}sp{\'a}l}, {Kostrzewa-Rutkowska}, {Kruszy{\'n}ska}, {Kun}, {Laizeau}, {Lambert}, {Lanza}, {Lasne}, {Le Campion}, {Lebreton}, {Lebzelter}, {Leccia}, {Leclerc}, {Lecoeur-Taibi}, {Liao}, {Licata}, {Lindstr{\o}m}, {Lister}, {Livanou}, {Lobel}, {Lorca}, {Loup}, {Madrero Pardo}, {Magdaleno Romeo}, {Managau}, {Mann}, {Manteiga}, {Marchant}, {Marconi}, {Marcos}, {Marcos Santos}, {Mar{\'\i}n Pina}, {Marinoni}, {Marocco}, {Marshall}, {Martin Polo}, {Mart{\'\i}n-Fleitas}, {Marton}, {Mary}, {Masip}, {Massari}, {Mastrobuono-Battisti}, {Mazeh}, {McMillan}, {Messina}, {Michalik},
  {Millar}, {Mints}, {Molina}, {Molinaro}, {Moln{\'a}r}, {Monari}, {Mongui{\'o}}, {Montegriffo}, {Montero}, {Mor}, {Mora}, {Morbidelli}, {Morel}, {Morris}, {Muraveva}, {Murphy}, {Musella}, {Nagy}, {Noval}, {Oca{\~n}a}, {Ogden}, {Ordenovic}, {Osinde}, {Pagani}, {Pagano}, {Palaversa}, {Palicio}, {Pallas-Quintela}, {Panahi}, {Payne-Wardenaar}, {Pe{\~n}alosa Esteller}, {Penttil{\"a}}, {Pichon}, {Piersimoni}, {Pineau}, {Plachy}, {Plum}, {Poggio}, {Pr{\v{s}}a}, {Pulone}, {Racero}, {Ragaini}, {Rainer}, {Raiteri}, {Rambaux}, {Ramos}, {Ramos-Lerate}, {Re Fiorentin}, {Regibo}, {Richards}, {Rios Diaz}, {Ripepi}, {Riva}, {Rix}, {Rixon}, {Robichon}, {Robin}, {Robin}, {Roelens}, {Rogues}, {Rohrbasser}, {Romero-G{\'o}mez}, {Rowell}, {Royer}, {Ruz Mieres}, {Rybicki}, {Sadowski}, {S{\'a}ez N{\'u}{\~n}ez}, {Sagrist{\`a} Sell{\'e}s}, {Sahlmann}, {Salguero}, {Samaras}, {Sanchez Gimenez}, {Sanna}, {Santove{\~n}a}, {Sarasso}, {Schultheis}, {Sciacca}, {Segol}, {Segovia}, {S{\'e}gransan}, {Semeux}, {Shahaf}, {Siddiqui}, {Siebert},
  {Siltala}, {Silvelo}, {Slezak}, {Slezak}, {Smart}, {Snaith}, {Solano}, {Solitro}, {Souami}, {Souchay}, {Spagna}, {Spina}, {Spoto}, {Steele}, {Steidelm{\"u}ller}, {Stephenson}, {S{\"u}veges}, {Surdej}, {Szabados}, {Szegedi-Elek}, {Taris}, {Taylor}, {Teixeira}, {Tolomei}, {Tonello}, {Torra}, {Torra}, {Torralba Elipe}, {Trabucchi}, {Tsounis}, {Turon}, {Ulla}, {Unger}, {Vaillant}, {van Dillen}, {van Reeven}, {Vanel}, {Vecchiato}, {Viala}, {Vicente}, {Voutsinas}, {Weiler}, {Wevers}, {Wyrzykowski}, {Yoldas}, {Yvard}, {Zhao}, {Zorec}, {Zucker}, \& {Zwitter}}]{GaiaDR3}
{Gaia Collaboration}, {Vallenari}, A., {Brown}, A.~G.~A., {et~al.} 2023, \aap, 674, A1, \dodoi{10.1051/0004-6361/202243940}

\bibitem[{{GRAVITY Collaboration} {et~al.}(2019){GRAVITY Collaboration}, {Abuter}, {Amorim}, {Baub{\"o}ck}, {Berger}, {Bonnet}, {Brandner}, {Cl{\'e}net}, {Coud{\'e} Du Foresto}, {de Zeeuw}, {Dexter}, {Duvert}, {Eckart}, {Eisenhauer}, {F{\"o}rster Schreiber}, {Garcia}, {Gao}, {Gendron}, {Genzel}, {Gerhard}, {Gillessen}, {Habibi}, {Haubois}, {Henning}, {Hippler}, {Horrobin}, {Jim{\'e}nez-Rosales}, {Jocou}, {Kervella}, {Lacour}, {Lapeyr{\`e}re}, {Le Bouquin}, {L{\'e}na}, {Ott}, {Paumard}, {Perraut}, {Perrin}, {Pfuhl}, {Rabien}, {Rodriguez Coira}, {Rousset}, {Scheithauer}, {Sternberg}, {Straub}, {Straubmeier}, {Sturm}, {Tacconi}, {Vincent}, {von Fellenberg}, {Waisberg}, {Widmann}, {Wieprecht}, {Wiezorrek}, {Woillez}, \& {Yazici}}]{Gravity2019}
{GRAVITY Collaboration}, {Abuter}, R., {Amorim}, A., {et~al.} 2019, \aap, 625, L10, \dodoi{10.1051/0004-6361/201935656}

\bibitem[{Harris {et~al.}(2020{\natexlab{a}})Harris, Millman, van~der Walt, Gommers, Virtanen, Cournapeau, Wieser, Taylor, Berg, Smith, Kern, Picus, Hoyer, van Kerkwijk, Brett, Haldane, del R{\'{i}}o, Wiebe, Peterson, G{\'{e}}rard-Marchant, Sheppard, Reddy, Weckesser, Abbasi, Gohlke, \& Oliphant}]{Numpy}
Harris, C.~R., Millman, K.~J., van~der Walt, S.~J., {et~al.} 2020{\natexlab{a}}, Nature, 585, 357, \dodoi{10.1038/s41586-020-2649-2}

\bibitem[{Harris {et~al.}(2020{\natexlab{b}})Harris, Millman, van~der Walt, Gommers, Virtanen, Cournapeau, Wieser, Taylor, Berg, Smith, Kern, Picus, Hoyer, van Kerkwijk, Brett, Haldane, del R{\'{i}}o, Wiebe, Peterson, G{\'{e}}rard-Marchant, Sheppard, Reddy, Weckesser, Abbasi, Gohlke, \& Oliphant}]{Numpy2020}
---. 2020{\natexlab{b}}, Nature, 585, 357, \dodoi{10.1038/s41586-020-2649-2}

\bibitem[{{Hasselquist} {et~al.}(2021){Hasselquist}, {Hayes}, {Lian}, {Weinberg}, {Zasowski}, {Horta}, {Beaton}, {Feuillet}, {Garro}, {Gallart}, {Smith}, {Holtzman}, {Minniti}, {Lacerna}, {Shetrone}, {J{\"o}nsson}, {Cioni}, {Fillingham}, {Cunha}, {O'Connell}, {Fern{\'a}ndez-Trincado}, {Mu{\~n}oz}, {Schiavon}, {Almeida}, {Anguiano}, {Beers}, {Bizyaev}, {Brownstein}, {Cohen}, {Frinchaboy}, {Garc{\'\i}a-Hern{\'a}ndez}, {Geisler}, {Lane}, {Majewski}, {Nidever}, {Nitschelm}, {Povick}, {Price-Whelan}, {Roman-Lopes}, {Rosado}, {Sobeck}, {Stringfellow}, {Valenzuela}, {Villanova}, \& {Vincenzo}}]{Hasselquist2021}
{Hasselquist}, S., {Hayes}, C.~R., {Lian}, J., {et~al.} 2021, \apj, 923, 172, \dodoi{10.3847/1538-4357/ac25f9}

\bibitem[{{Hawkins} {et~al.}(2015){Hawkins}, {Jofr{\'e}}, {Masseron}, \& {Gilmore}}]{Hawkins2015}
{Hawkins}, K., {Jofr{\'e}}, P., {Masseron}, T., \& {Gilmore}, G. 2015, \mnras, 453, 758, \dodoi{10.1093/mnras/stv1586}

\bibitem[{{Haywood} {et~al.}(2018){Haywood}, {Di Matteo}, {Lehnert}, {Snaith}, {Khoperskov}, \& {G{\'o}mez}}]{Haywood2018}
{Haywood}, M., {Di Matteo}, P., {Lehnert}, M.~D., {et~al.} 2018, \apj, 863, 113, \dodoi{10.3847/1538-4357/aad235}

\bibitem[{{Helmi} {et~al.}(2018){Helmi}, {Babusiaux}, {Koppelman}, {Massari}, {Veljanoski}, \& {Brown}}]{Helmi2018}
{Helmi}, A., {Babusiaux}, C., {Koppelman}, H.~H., {et~al.} 2018, \nat, 563, 85, \dodoi{10.1038/s41586-018-0625-x}

\bibitem[{{Helmi} {et~al.}(1999){Helmi}, {White}, {de Zeeuw}, \& {Zhao}}]{Helmi_Streams}
{Helmi}, A., {White}, S. D.~M., {de Zeeuw}, P.~T., \& {Zhao}, H. 1999, \nat, 402, 53, \dodoi{10.1038/46980}

\bibitem[{{Horta} {et~al.}(2022){Horta}, {Ness}, {Rybizki}, {Schiavon}, \& {Buder}}]{Horta2022}
{Horta}, D., {Ness}, M.~K., {Rybizki}, J., {Schiavon}, R.~P., \& {Buder}, S. 2022, \mnras, 513, 5477, \dodoi{10.1093/mnras/stac953}

\bibitem[{{Horta} {et~al.}(2020){Horta}, {Schiavon}, {Mackereth}, {Beers}, {Fern{\'a}ndez-Trincado}, {Frinchaboy}, {Garc{\'\i}a-Hern{\'a}ndez}, {Geisler}, {Hasselquist}, {J{\"o}nsson}, {Lane}, {Majewski}, {M{\'e}sz{\'a}ros}, {Bidin}, {Nataf}, {Roman-Lopes}, {Nitschelm}, {Vargas-Gonz{\'a}lez}, \& {Zasowski}}]{Horta2020}
{Horta}, D., {Schiavon}, R.~P., {Mackereth}, J.~T., {et~al.} 2020, \mnras, 493, 3363, \dodoi{10.1093/mnras/staa478}

\bibitem[{{Horta} {et~al.}(2021){Horta}, {Schiavon}, {Mackereth}, {Pfeffer}, {Mason}, {Kisku}, {Fragkoudi}, {Allende Prieto}, {Cunha}, {Hasselquist}, {Holtzman}, {Majewski}, {Nataf}, {O'Connell}, {Schultheis}, \& {Smith}}]{Horta2021}
---. 2021, \mnras, 500, 1385, \dodoi{10.1093/mnras/staa2987}

\bibitem[{{Horta} {et~al.}(2023){Horta}, {Schiavon}, {Mackereth}, {Weinberg}, {Hasselquist}, {Feuillet}, {O'Connell}, {Anguiano}, {Allende-Prieto}, {Beaton}, {Bizyaev}, {Cunha}, {Geisler}, {Garc{\'\i}a-Hern{\'a}ndez}, {Holtzman}, {J{\"o}nsson}, {Lane}, {Majewski}, {M{\'e}sz{\'a}ros}, {Minniti}, {Nitschelm}, {Shetrone}, {Smith}, \& {Zasowski}}]{Horta2023}
---. 2023, \mnras, 520, 5671, \dodoi{10.1093/mnras/stac3179}

\bibitem[{Hunter(2007{\natexlab{a}})}]{Matplotlib}
Hunter, J.~D. 2007{\natexlab{a}}, Computing in Science \& Engineering, 9, 90, \dodoi{10.1109/MCSE.2007.55}

\bibitem[{Hunter(2007{\natexlab{b}})}]{Matplotlib2007}
---. 2007{\natexlab{b}}, Computing in Science \& Engineering, 9, 90, \dodoi{10.1109/MCSE.2007.55}

\bibitem[{{Ibata} {et~al.}(1994){Ibata}, {Gilmore}, \& {Irwin}}]{Ibata1994}
{Ibata}, R.~A., {Gilmore}, G., \& {Irwin}, M.~J. 1994, \nat, 370, 194, \dodoi{10.1038/370194a0}

\bibitem[{{Iwamoto} {et~al.}(1999){Iwamoto}, {Brachwitz}, {Nomoto}, {Kishimoto}, {Umeda}, {Hix}, \& {Thielemann}}]{Iwamoto1999}
{Iwamoto}, K., {Brachwitz}, F., {Nomoto}, K., {et~al.} 1999, \apjs, 125, 439, \dodoi{10.1086/313278}

\bibitem[{{Karakas}(2010)}]{Karakas2010}
{Karakas}, A.~I. 2010, \mnras, 403, 1413, \dodoi{10.1111/j.1365-2966.2009.16198.x}

\bibitem[{{Kennicutt}(1989)}]{Kennicutt1989}
{Kennicutt}, R. 1989, \apj, 344, 685, \dodoi{10.1086/167834}

\bibitem[{{Kobayashi} {et~al.}(2006){Kobayashi}, {Umeda}, {Nomoto}, {Tominaga}, \& {Ohkubo}}]{Kobayashi2006}
{Kobayashi}, C., {Umeda}, H., {Nomoto}, K., {Tominaga}, N., \& {Ohkubo}, T. 2006, \apj, 653, 1145, \dodoi{10.1086/508914}

\bibitem[{{Koppelman} {et~al.}(2019){Koppelman}, {Helmi}, {Massari}, {Price-Whelan}, \& {Starkenburg}}]{Koppelman2019}
{Koppelman}, H.~H., {Helmi}, A., {Massari}, D., {Price-Whelan}, A.~M., \& {Starkenburg}, T.~K. 2019, \aap, 631, L9, \dodoi{10.1051/0004-6361/201936738}

\bibitem[{{Kroupa} {et~al.}(1993){Kroupa}, {Tout}, \& {Gilmore}}]{Kroupa1993}
{Kroupa}, P., {Tout}, C.~A., \& {Gilmore}, G. 1993, \mnras, 262, 545, \dodoi{10.1093/mnras/262.3.545}

\bibitem[{{Kruijssen} {et~al.}(2019){Kruijssen}, {Pfeffer}, {Reina-Campos}, {Crain}, \& {Bastian}}]{Kruijssen2019}
{Kruijssen}, J.~M.~D., {Pfeffer}, J.~L., {Reina-Campos}, M., {Crain}, R.~A., \& {Bastian}, N. 2019, \mnras, 486, 3180, \dodoi{10.1093/mnras/sty1609}

\bibitem[{{Lane} {et~al.}(2023){Lane}, {Bovy}, \& {Mackereth}}]{Lane2023}
{Lane}, J. M.~M., {Bovy}, J., \& {Mackereth}, J.~T. 2023, \mnras, 526, 1209, \dodoi{10.1093/mnras/stad2834}

\bibitem[{Limberg {et~al.}(2022)Limberg, Souza, P{\' e}rez-Villegas, Rossi, Perottoni, \& Santucci}]{Limberg2022}
Limberg, G., Souza, S.~O., P{\' e}rez-Villegas, A., {et~al.} 2022, The Astrophysical Journal, 935, 109, \dodoi{10.3847/1538-4357/ac8159}

\bibitem[{{Mackereth} {et~al.}(2019){Mackereth}, {Schiavon}, {Pfeffer}, {Hayes}, {Bovy}, {Anguiano}, {Allende Prieto}, {Hasselquist}, {Holtzman}, {Johnson}, {Majewski}, {O'Connell}, {Shetrone}, {Tissera}, \& {Fern{\'a}ndez-Trincado}}]{Mackereth2019}
{Mackereth}, J.~T., {Schiavon}, R.~P., {Pfeffer}, J., {et~al.} 2019, \mnras, 482, 3426, \dodoi{10.1093/mnras/sty2955}

\bibitem[{{Majewski} {et~al.}(2017){Majewski}, {Schiavon}, {Frinchaboy}, {Allende Prieto}, {Barkhouser}, {Bizyaev}, {Blank}, {Brunner}, {Burton}, {Carrera}, {Chojnowski}, {Cunha}, {Epstein}, {Fitzgerald}, {Garc{\'\i}a P{\'e}rez}, {Hearty}, {Henderson}, {Holtzman}, {Johnson}, {Lam}, {Lawler}, {Maseman}, {M{\'e}sz{\'a}ros}, {Nelson}, {Nguyen}, {Nidever}, {Pinsonneault}, {Shetrone}, {Smee}, {Smith}, {Stolberg}, {Skrutskie}, {Walker}, {Wilson}, {Zasowski}, {Anders}, {Basu}, {Beland}, {Blanton}, {Bovy}, {Brownstein}, {Carlberg}, {Chaplin}, {Chiappini}, {Eisenstein}, {Elsworth}, {Feuillet}, {Fleming}, {Galbraith-Frew}, {Garc{\'\i}a}, {Garc{\'\i}a-Hern{\'a}ndez}, {Gillespie}, {Girardi}, {Gunn}, {Hasselquist}, {Hayden}, {Hekker}, {Ivans}, {Kinemuchi}, {Klaene}, {Mahadevan}, {Mathur}, {Mosser}, {Muna}, {Munn}, {Nichol}, {O'Connell}, {Parejko}, {Robin}, {Rocha-Pinto}, {Schultheis}, {Serenelli}, {Shane}, {Silva Aguirre}, {Sobeck}, {Thompson}, {Troup}, {Weinberg}, \& {Zamora}}]{Apogee}
{Majewski}, S.~R., {Schiavon}, R.~P., {Frinchaboy}, P.~M., {et~al.} 2017, \aj, 154, 94, \dodoi{10.3847/1538-3881/aa784d}

\bibitem[{{Malhan} {et~al.}(2022){Malhan}, {Ibata}, {Sharma}, {Famaey}, {Bellazzini}, {Carlberg}, {D'Souza}, {Yuan}, {Martin}, \& {Thomas}}]{Malhan2022}
{Malhan}, K., {Ibata}, R.~A., {Sharma}, S., {et~al.} 2022, \apj, 926, 107, \dodoi{10.3847/1538-4357/ac4d2a}

\bibitem[{{Maoz} {et~al.}(2014){Maoz}, {Mannucci}, \& {Nelemans}}]{Maoz2014}
{Maoz}, D., {Mannucci}, F., \& {Nelemans}, G. 2014, \araa, 52, 107, \dodoi{10.1146/annurev-astro-082812-141031}

\bibitem[{{Matsuno} {et~al.}(2019){Matsuno}, {Aoki}, \& {Suda}}]{Matsuno2019}
{Matsuno}, T., {Aoki}, W., \& {Suda}, T. 2019, \apjl, 874, L35, \dodoi{10.3847/2041-8213/ab0ec0}

\bibitem[{{McMillan}(2017)}]{McMillan2017}
{McMillan}, P.~J. 2017, \mnras, 465, 76, \dodoi{10.1093/mnras/stw2759}

\bibitem[{{Mori} {et~al.}(2024){Mori}, {Di Matteo}, {Salvadori}, {Khoperskov}, {Pagnini}, \& {Haywood}}]{Mori2024}
{Mori}, A., {Di Matteo}, P., {Salvadori}, S., {et~al.} 2024, \aap, 690, A136, \dodoi{10.1051/0004-6361/202449291}

\bibitem[{{Myeong} {et~al.}(2022){Myeong}, {Belokurov}, {Aguado}, {Evans}, {Caldwell}, \& {Bradley}}]{Myeong2022}
{Myeong}, G.~C., {Belokurov}, V., {Aguado}, D.~S., {et~al.} 2022, \apj, 938, 21, \dodoi{10.3847/1538-4357/ac8d68}

\bibitem[{{Myeong} {et~al.}(2019){Myeong}, {Vasiliev}, {Iorio}, {Evans}, \& {Belokurov}}]{Myeong2019}
{Myeong}, G.~C., {Vasiliev}, E., {Iorio}, G., {Evans}, N.~W., \& {Belokurov}, V. 2019, \mnras, 488, 1235, \dodoi{10.1093/mnras/stz1770}

\bibitem[{{Naidu} {et~al.}(2020){Naidu}, {Conroy}, {Bonaca}, {Johnson}, {Ting}, {Caldwell}, {Zaritsky}, \& {Cargile}}]{Naidu2020}
{Naidu}, R.~P., {Conroy}, C., {Bonaca}, A., {et~al.} 2020, \apj, 901, 48, \dodoi{10.3847/1538-4357/abaef4}

\bibitem[{{Nissen} {et~al.}(2024){Nissen}, {Amarsi}, {Sk{\'u}lad{\'o}ttir}, \& {Schuster}}]{Nissen2024}
{Nissen}, P.~E., {Amarsi}, A.~M., {Sk{\'u}lad{\'o}ttir}, {\'A}., \& {Schuster}, W.~J. 2024, \aap, 682, A116, \dodoi{10.1051/0004-6361/202348392}

\bibitem[{{Nissen} \& {Schuster}(1997)}]{NissenSchuster1997}
{Nissen}, P.~E., \& {Schuster}, W.~J. 1997, \aap, 326, 751

\bibitem[{Pedregosa {et~al.}(2011)Pedregosa, Varoquaux, Gramfort, Michel, Thirion, Grisel, Blondel, Prettenhofer, Weiss, Dubourg, Vanderplas, Passos, Cournapeau, Brucher, Perrot, \& Duchesnay}]{scikit-learn}
Pedregosa, F., Varoquaux, G., Gramfort, A., {et~al.} 2011, Journal of Machine Learning Research, 12, 2825

\bibitem[{{Pepe} {et~al.}(2025){Pepe}, {Palla}, {Matteucci}, \& {Spitoni}}]{Pepe2025}
{Pepe}, E., {Palla}, M., {Matteucci}, F., \& {Spitoni}, E. 2025, \aap, 694, A19, \dodoi{10.1051/0004-6361/202452698}

\bibitem[{{Schiavon} {et~al.}(2024){Schiavon}, {Phillips}, {Myers}, {Horta}, {Minniti}, {Allende Prieto}, {Anguiano}, {Beaton}, {Beers}, {Brownstein}, {Cohen}, {Fern{\'a}ndez-Trincado}, {Frinchaboy}, {J{\"o}nsson}, {Kisku}, {Lane}, {Majewski}, {Mason}, {M{\'e}sz{\'a}ros}, \& {Stringfellow}}]{apogee_gc}
{Schiavon}, R.~P., {Phillips}, S.~G., {Myers}, N., {et~al.} 2024, \mnras, 528, 1393, \dodoi{10.1093/mnras/stad3020}

\bibitem[{{Schmidt}(1959)}]{Schmidt1959}
{Schmidt}, M. 1959, \apj, 129, 243, \dodoi{10.1086/146614}

\bibitem[{{Sch{\"o}nrich} {et~al.}(2010){Sch{\"o}nrich}, {Binney}, \& {Dehnen}}]{Schonrich2010}
{Sch{\"o}nrich}, R., {Binney}, J., \& {Dehnen}, W. 2010, \mnras, 403, 1829, \dodoi{10.1111/j.1365-2966.2010.16253.x}

\bibitem[{{Spitoni} {et~al.}(2020){Spitoni}, {Verma}, {Silva Aguirre}, \& {Calura}}]{Spitonietal2020}
{Spitoni}, E., {Verma}, K., {Silva Aguirre}, V., \& {Calura}, F. 2020, \aap, 635, A58, \dodoi{10.1051/0004-6361/201937275}

\bibitem[{{Spitoni} {et~al.}(2021){Spitoni}, {Verma}, {Silva Aguirre}, {Vincenzo}, {Matteucci}, {Vai{\v{c}}ekauskait{\.{e}}}, {Palla}, {Grisoni}, \& {Calura}}]{Spitonietal2021}
{Spitoni}, E., {Verma}, K., {Silva Aguirre}, V., {et~al.} 2021, \aap, 647, A73, \dodoi{10.1051/0004-6361/202039864}

\bibitem[{{Steinmetz} {et~al.}(2006){Steinmetz}, {Zwitter}, {Siebert}, {Watson}, {Freeman}, {Munari}, {Campbell}, {Williams}, {Seabroke}, {Wyse}, {Parker}, {Bienaym{\'e}}, {Roeser}, {Gibson}, {Gilmore}, {Grebel}, {Helmi}, {Navarro}, {Burton}, {Cass}, {Dawe}, {Fiegert}, {Hartley}, {Russell}, {Saunders}, {Enke}, {Bailin}, {Binney}, {Bland-Hawthorn}, {Boeche}, {Dehnen}, {Eisenstein}, {Evans}, {Fiorucci}, {Fulbright}, {Gerhard}, {Jauregi}, {Kelz}, {Mijovi{\'c}}, {Minchev}, {Parmentier}, {Pe{\~n}arrubia}, {Quillen}, {Read}, {Ruchti}, {Scholz}, {Siviero}, {Smith}, {Sordo}, {Veltz}, {Vidrih}, {von Berlepsch}, {Boyle}, \& {Schilbach}}]{Steinmetzetal2006}
{Steinmetz}, M., {Zwitter}, T., {Siebert}, A., {et~al.} 2006, \aj, 132, 1645, \dodoi{10.1086/506564}

\bibitem[{Van~Rossum \& Drake~Jr(1995)}]{Python}
Van~Rossum, G., \& Drake~Jr, F.~L. 1995, Python reference manual (Centrum voor Wiskunde en Informatica Amsterdam)

\bibitem[{{Vandenberg}(1983)}]{tsne}
{Vandenberg}, D.~A. 1983, \apjs, 51, 29, \dodoi{10.1086/190839}

\bibitem[{Vincenzo {et~al.}(2019)Vincenzo, Spitoni, Calura, Matteucci, Silva~Aguirre, Miglio, \& Cescutti}]{Vincenzo2019}
Vincenzo, F., Spitoni, E., Calura, F., {et~al.} 2019, Monthly Notices of the Royal Astronomical Society: Letters, 487, L47, \dodoi{10.1093/mnrasl/slz070}

\bibitem[{{Yuan} {et~al.}(2020){Yuan}, {Chang}, {Beers}, \& {Huang}}]{Yuan2020}
{Yuan}, Z., {Chang}, J., {Beers}, T.~C., \& {Huang}, Y. 2020, \apjl, 898, L37, \dodoi{10.3847/2041-8213/aba49f}

\end{thebibliography}
\bibliographystyle{aasjournal}



\end{document}